\documentclass{mn2e}
\usepackage{amssymb,graphicx}

\def\LCDM{{$\Lambda$CDM}}
\def\Mstar{{M$^*$}}

\def\Mpc{{\,h^{-1}\,{\rm Mpc}}} 
\def\Mpccube{{\,h^{-3}\,{\rm Mpc^3}}} 
\def\Gpccube{{\,h^{-3}\,{\rm Gpc^3}}} 

\def\etal{{et~al.}}

\def\nsub{{${\rm N_{\rm sub}}$}}

\newcommand{\plotfull}[1] 
           {\centering \leavevmode 
             \includegraphics[width=\textwidth,clip=]{#1}}
\newcommand{\plotone}[1] 
           {\centering \leavevmode 
             \includegraphics[width=\columnwidth,clip=]{#1}} 
\newcommand{\plottwo}[2] 
           {\centering \leavevmode 
             \includegraphics[width=\columnwidth,clip=]{#1}
             \hfill 
             \includegraphics[width=\columnwidth,clip=]{#2}}

\def\simlt{\lower.5ex\hbox{$\; \buildrel < \over \sim \;$}}
\def\simgt{\lower.5ex\hbox{$\; \buildrel > \over \sim \;$}}

\title[SAGS-IV: An objective way to quantify the impact of LSS]
{Statistical Analysis of Galaxy Surveys-IV: \\
An objective way to quantify 
the impact of superstructures on galaxy clustering statistics}

\author[P. Norberg, E. Gazta\~naga, C.M. Baugh \& D.J. Croton]{ 
P. Norberg$^1$, 
E. Gazta\~{n}aga$^2$, 
C. M. Baugh$^3$, 
D. J. Croton$^{4}$\\ 
$^1$SUPA\thanks{The Scottish Universities Physics Alliance}, Institute for Astronomy, University of Edinburgh, Royal Observatory, Blackford Hill, Edinburgh, EH9 3HJ, UK.\\
$^2$Instituto de Ciencias del Espacio (IEEC/CSIC), F. de Ciencias UAB, Torre C5- Par-2a, Bellaterra, 08193 Barcelona, Spain.\\
$^3$Institute for Computational Cosmology, Department of Physics,
University of Durham, South Road, Durham DH1 3LE, UK.\\
$^4$Centre for Astrophysics and Supercomputing, Swinburne University of Technology, Mail H39, PO Box 218, Hawthorn, Victoria, \\
3122, Australia\\
} 
 
\date{Accepted ---. Received ---;in original form ---}

\begin{document} 
 
\maketitle 
 
%\label{firstpage} 

\begin{abstract} 
For galaxy clustering to provide robust constraints on 
cosmological parameters and galaxy formation models, it 
is essential to make reliable estimates of the errors 
on clustering measurements. We present a new technique, 
based on a spatial Jackknife (JK) resampling, which provides 
an objective way to estimate errors on clustering statistics. 
Our approach allows us to set the appropriate size for 
the Jackknife subsamples. The method also provides a means 
to assess the impact of individual regions on the measured 
clustering, and thereby to establish whether or not a given 
galaxy catalogue is dominated by one or several large structures, 
preventing it to be considered as a ``fair sample''.  
We apply this methodology to the two- and three-point correlation 
functions measured from a volume limited sample of \Mstar\ 
galaxies drawn from data release seven of the Sloan Digital Sky 
Survey (SDSS). 
The frequency of jackknife subsample outliers in the data is shown to
be consistent 
with that seen in large N-body simulations of clustering in the 
cosmological constant plus cold dark matter cosmology. 
We also present a comparison of the three-point correlation 
function in SDSS and 2dFGRS using this approach and
find consistent measurements between the two samples.
\end{abstract} 
                                   

\section{Introduction}

The clustering of galaxies has the potential to place constraints on 
the values of fundamental cosmological parameters and to probe the 
efficiency of galaxy formation in dark matter haloes of different 
masses. A key assumption made when interpreting clustering measurements   
is that the sample used is representative of a much larger volume of the 
Universe. The hope is that the survey volume is sufficiently large that 
the clustering measurements made from it should agree with the mean of 
measurements obtained from an ensemble of similar volumes (which is, 
in general, not feasible to carry out). This is the ``fair sample'' 
hypothesis (Peebles 1980).

This ``fair sample'' hypothesis, commonly invoked in large scale
structure analyses, is often abused in the literature. The
hypothesis relies on two conditions: (a) that the clustering statistic
for one survey is not biased with respect
to the mean measurement from an ensemble of independent but similar
surveys; (b) that the error
estimate on the statistic is properly characterized, i.e. it accounts
for the variance seen in the ensemble of
measurements (that is most often not achievable with real data). The
latter point is often ignored and
samples are referred to as ``unfair'' when the clustering statistic
(with its associated errors) is at odds with either
that from other samples (and their errors) or with the expected,
theoretical value. The point we want to stress in
this paper is that most surveys are likely to be ``fair'' but that the
associated error analysis is likely to be
``unfair''. This problem arises because a simplistic (or computationally
inexpensive) approach to estimating the errors
has been implemented and insufficient attention has been given to the
inherent limitations of the method used. The
recent extensive use of covariance matrices to account for the full
error budget has brushed aside the important
discussion of the limitations of the error method considered (as
discussed at length in e.g. Norberg \etal\ 2009 for 2-point clustering
statistics).

Surveys of the local Universe have increased in size by several orders 
of magnitude since the late 1990s, with over one million galaxy 
redshifts measured to date, spanning volumes running to several 
hundreds of cubic megaparsecs (York \etal\ 2000; Colless \etal\ 2001, 2003).   
Nevertheless, some clustering analyses have thrown up unusual results 
which suggest that current surveys may not, at least under 
certain conditions, be fair samples of the Universe. These anomalous 
results were first revealed in the form of the hierarchical amplitudes, 
the ratio of higher order correlation functions to a power of the 
two-point function, measured from volume limited samples drawn from 
the two-degree field galaxy redshift survey (2dFGRS; Baugh \etal\ 2004; 
Croton \etal\ 2004; Gaztanaga \etal\ 2005). 
The hierarchical amplitudes measured from the 
2dFGRS displayed an upturn on large scales which is not expected in 
the gravitational instability scenario (e.g.\ Baugh, Gaztanaga \& 
Efstathiou 1995). Croton \etal\ (2004) demonstrated that this was due 
to two particularly ``hot'' cells in the cell count distribution, which 
contained unusually large numbers of galaxies. On blanking out these cells 
from the survey and removing them from the count probability distribution, 
the hierarchical amplitudes resembled the theoretical expectations more 
closely. 
Similar conclusions were reached in separate analyses of the two-point
and three point correlation functions measured from the Sloan Digital
Sky Survey (SDSS; Zehavi \etal\ 2002, 2004, 2005, 2011; 
Nichol \etal\ 2006; McBride et al. 2011a).

One of the ``superstructures'' identified in the 2dFGRS analyses is
part of the SDSS ``Great Wall'' (Gott \etal\ 2005). Clustering
measurements from the magnitude limited 2dFGRS do not show any
significant  
change 
when this  
region is omitted (e.g.\ Cole \etal\ 2005). The influence of this structure over 
clustering measurments seems to depend on how the sample is constructed, 
an issue which we investigate further in this paper. 
The two point clustering analyses of Zehavi \etal\ (2002, 2005, 2011) on
SDSS have shown explicitly the influence of the SDSS ``Great Wall'' on
their measurments, with 
volume limited samples of \Mstar\ galaxies being particularly affected, whereas 
samples corresponding to brighter galaxy luminosities are less sensitive to 
the presence of this structure. Different volume limited samples will weight 
the structure differently, as it will be traced out by a different number of 
galaxies. Also, the volume of the sample changes when the luminosity 
defining it is varied; in a larger volume, other, similar structures may 
feature, diluting the impact of any single structure on clustering 
measurements. 

Croton \etal\ (2004, 2007), as well as Gazta\~naga et al (2005),
presented measurements with and without the galaxies in the hot
regions identified in their counts in cells analysis of the 2dFGRS,
with the simple aim of showing the contribution of this structure 
on the hierarchical moments. Their approach was specific to the 
counts in cells clustering 
analysis. In this paper we present an objective technique which can be applied 
to any clustering statistic. The new method we describe is a development 
of the Jackknife technique for error estimation (see e.g.\ Norberg
\etal\ 2009 for a review and application of this method and others to
galaxy clustering error estimations;
see also Zehavi \etal\ 2002 who gives the first comprehensive
description of the Jackknife method for clustering statistics, which
was used in the eariler clustering analysis of 
Roche \etal\ 1993 and Croft \etal\ 1999).  
Our approach can be used to assess whether or not a clustering signal is 
unduly influenced by the galaxy distribution in one particular region 
of a survey. Furthermore, we discuss a new statistic, the JK ensemble
fluctuation, which provides a robust assessment of such fluctuations in
the galaxy distribution. 
We note here 
%here\footnote{We thank the referee for pointing this out.}
that McBride \etal\ (2011a) present 
%in their section 4.5 a very 
a similar analysis for their SDSS 3-point function measurement (in 
particular their Figs. 11, 12, and 13 are very relevant for the
present paper). The main difference between their work and ours is that
we take the study of the influence of extreme structure significantly
further, by introducing a new statistic the JK ensemble fluctuation.

This paper is laid out as follows. In Section~\ref{sec:data} we 
present the galaxy survey data used, which is the seventh data release 
of the SDSS, the associated completeness masks, the division of the SDSS 
into Jackknife regions (or JK quilts) and the clustering statistics used. 
Section~\ref{sec:jk_outliers} is devoted to our new objective
method based on Jackknife resampling of the data, which is applied to the
\Mstar\ SDSS sample, and shows the impact of large structures or
``superstructures'' on 2 and 3-point statistics in redshift
space\footnote{This paper considers only analyses in redshift space,
  even though our method is perfectly valid for other clustering
  statistics in both real and redshift space.}. 
A new statistic, the JK ensemble fluctuation, is introduced to
quantify the importance of outliers. 
In Section~\ref{sec:galform_impact} we illustrate the performance 
of this new statistic and show how it successfully identifies unusual 
regions in SDSS DR7 and sets their significance. 
In Section~\ref{sec:lcdm_sim} apply our Jackknife approach to 
\LCDM\ simulations to show that they display similar outliers 
to those seen in the data. In Section~\ref{sec:SDSS_2dFGRS} we revist
the 3-pt correlation function analysis of Gazta\~naga et al. (2005),
but this time using our new methodology based on the JK ensemble
fluctuation.
A summary is given in Section~\ref{sec:conclusion}.  
Throughout we assume a standard cosmological model, with 
$\Omega_{\rm M}=0.25$, $\Omega_{\Lambda}=0.75$ and a value for the
Hubble parameter of $h=H_{0}/(100\,{\rm km\,s}^{-1}{\rm Mpc}^{-1})$.

\begin{figure}
%High-res
%\plotone{figs_highres/ra_dec_boot_compl_mask_sdss_zoom_highres.pdf}
%Low-res
\plotone{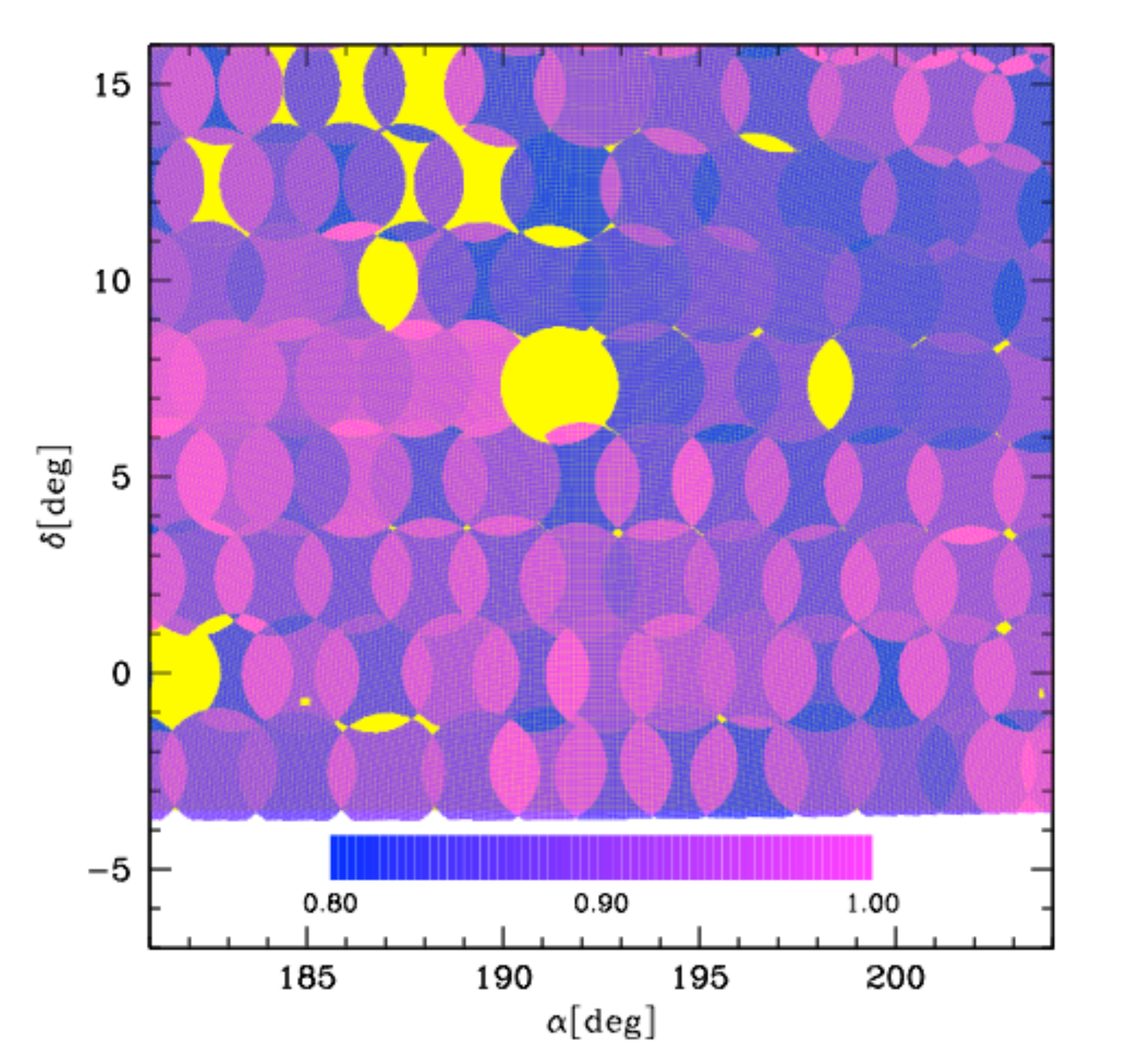}
\caption
{ 
A zoomed in view of the SDSS redshift completeness mask.
The colour bar indicates the redshift completeness fraction which
ranges from 0.8 to 1. Sectors (i.e.\ uniquely defined regions sampled
by the spectroscopic tiles) with completeness below 0.8 are shaded in
yellow and regions outside the survey area in white. 
}
\label{fig:mask_zoom}
\end{figure}

\section{Data and methodology}
\label{sec:data}

In this section we describe the galaxy data used (\S~2.1), the 
completeness mask of the survey (\S~2.2), the division of the 
survey into zones for Jackknife sampling (\S~2.3), the volume 
limited samples used in the clustering analysis presented in 
subsequent sections (\S~2.4) and the method estimation of 2 and 3 
point correlation functions (\S~2.5). 

\subsection{SDSS DR7 Data}

In this paper we use Data Release 7 (DR7) of the Sloan Digital 
Sky Survey (SDSS; Abazajian \etal\ 2009). DR7 covers 9380 deg$^2$ 
and contains 930\,000 galaxy redshifts. We select galaxies 
brighter than a Petrosian magnitude of $r<17.65$.
The magnitude limit is slightly brighter than that of the 
canonical main galaxy sample of $r=17.77$ (Strauss \etal\ 2002).   
We adopt this cut to avoid having to model a varying magnitude 
limit, which is needed to include all of the early SDSS data, since 
the magnitudes have been revised slightly since the early data 
were taken. 
In addition we also impose a bright magnitude limit of $r=15$, the
point at which SDSS galaxy magnitudes start to become less reliable.
The sample we consider contains 513k high quality, unique galaxy
redshifts with a median redshift of $z_{\rm med} \simeq 0.10$. The
precise choice of magnitude limit does not have an impact on our
results. In our analysis we use SDSS Petrosian magnitudes calibrated
using the prescription of Tucker \etal\ (2006) and corrected
for Galactic extinction.

\begin{figure*}
%High-res
%\plottwo{figs_highres/ra_dec_boot_quilt_ntot_25_highres.pdf}{figs_highres/ra_dec_boot_quilt_ntot_225_highres.pdf}
%Low-res
\plotfull{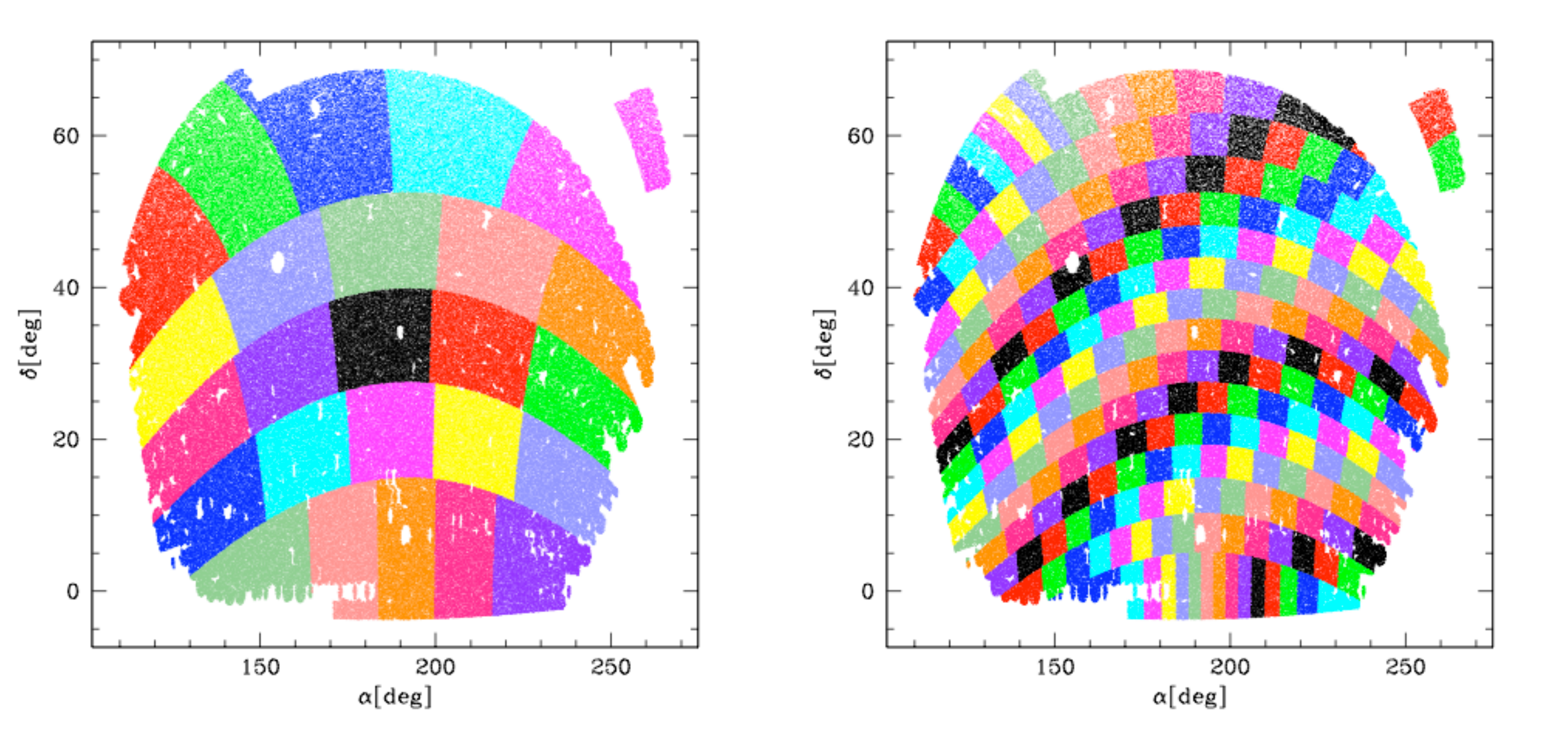}
\caption
{
{\it Left:} The SDSS survey divided into 5x5 Jackknife regions 
in the $\alpha - \delta$ plane. The 25 zones in the resulting 
%crazy quilt\footnote{http://en.wikipedia.org/wiki/Crazy_quilting}
crazy quilt have equal area after taking into account their 
spectroscopic completeness. 
{\it Right:} Same as the left-hand panel, but for the case of 225
Jackknife zones corresponding to a square 15x15 Jackknife quilt.  
}
\label{fig:sdss_quilt}
\end{figure*}

\subsection{SDSS survey mask}

In order to make robust and reliable estimates of clustering  
the incompleteness in the spectroscopic catalogue needs to be taken 
into account. We do this using a redshift completeness mask. The 
variation in the survey completeness on the sky is then used to modulate 
the density of unclustered or random points used in the estimation of 
the mean density in clustering analyses. 

We start by constructing a mask from the individual SDSS imaging scans. 
The imaging mask is pixelized using an equal-area projection, with a 
typical pixel area of $\sim 5.3$~arcmin$^2$. This is more than 
sufficient given the average number density of SDSS targets of $\sim 90$ 
galaxies deg$^{-2}$ and the typical scales we are interested in 
($s \ge 1 \Mpc$ at $z \ge 0.04$). Bad pixels, as labelled by 
the SDSS pipeline, are omitted from the imaging mask. 
%These include, for example, pixels affected by saturated stars. 
Next, using the position of the spectroscopic tiles, we create 
the spectroscopic completeness mask, following the method devised 
for the 2dFGRS survey (Colless \etal\ 2001; Norberg \etal\ 2002; 
Cole \etal\ 2005). The survey is divided up into sectors which 
are regions defined by the unique overlap of spectroscopic tiles. 
Any sector which contains fewer than 10 galaxies is merged with 
its most populous neighbouring sector. This ensures that 
the sector completeness is not affected by shot-noise, 
which would lead to a patchy redshift completeness mask. 
The redshift completeness in a sector is the ratio of galaxies 
with a measured (high quality) spectroscopic redshift divided by 
the total number of galaxies in the target catalogue in that sector.
The CasJobs SQL queries and data files used to retrieve the
information required to construct the SDSS survey masks (imaging and
spectroscopic) are given in Appendix~\ref{sec:appendix_query}. 

We show a small part of the survey mask we have constructed for the 
SDSS DR7 in Fig.~\ref{fig:mask_zoom}. Regions outside the survey
boundary are shaded white. Sectors which are less than 80\% complete
but which lie within the survey boundary are shaded yellow. The
remaining sectors have a spectroscopic completeness ranging between 80
and 100\%. It is these latter regions that are retained in our
subsequent clustering analysis. 

\begin{figure*}
\plotfull{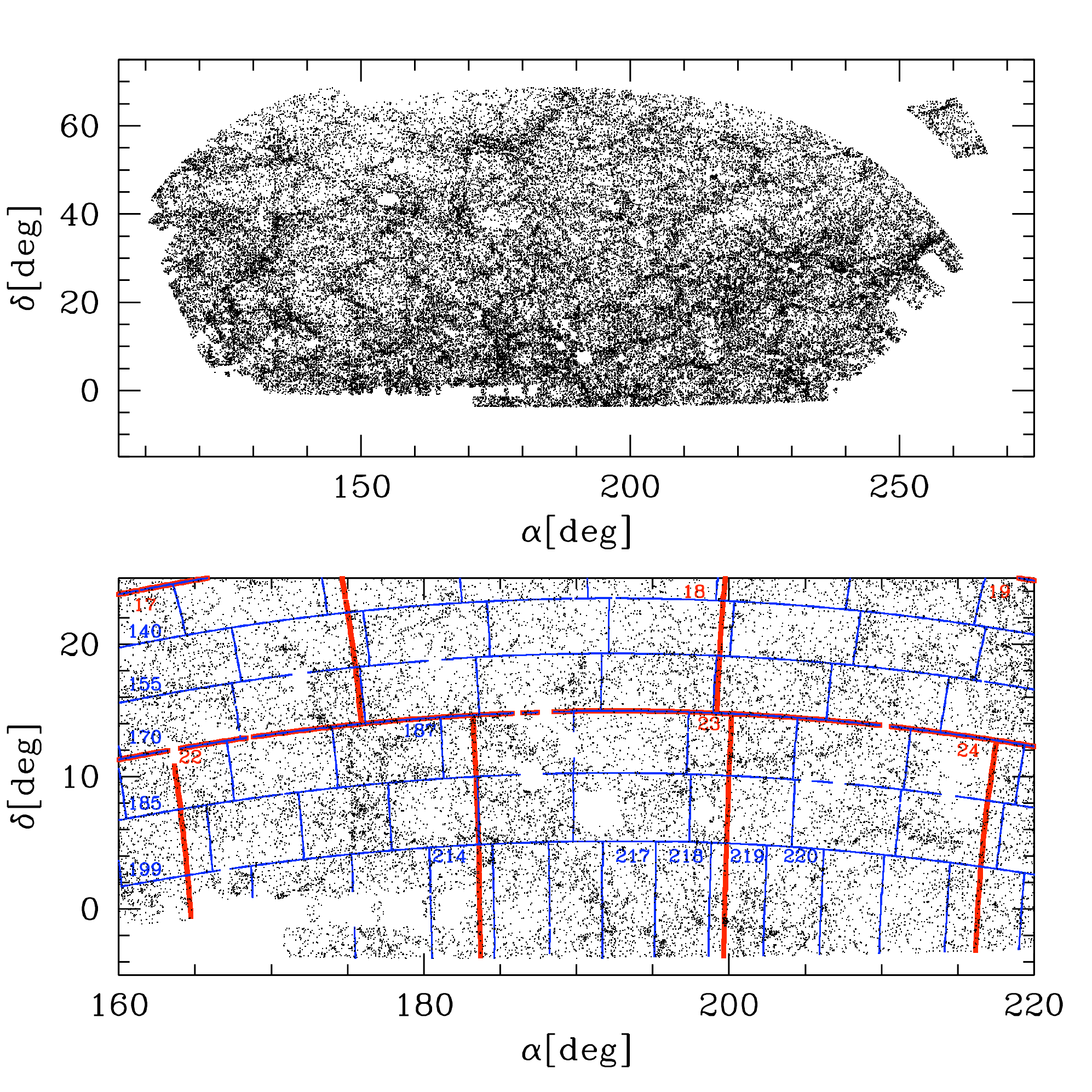}
\caption
{{\it Top:} The angular galaxy distribution of the {\tt ref}
volume limited sample (dots). 
%Overplotted (in yellow) is the mask corresponding to zone 23 (in the
%5x5 Jackknife quilt) which will be later identified as an outlier. 
{\it Bottom:}
A zoomed in view centered on zone 23. Overplotted are the 
boundaries of the zones for the 5x5 and 15x15 Jackknife quilts, 
shown in thick red and thin
blue linestyles respectively. The subzone numbers (colour coded 
according to the dimension of the quilt they belong to) 
are indicated for a limited number of
patches, some of which will be referred to in later sections.
The subzone number increases from left to right
from top to bottom. 
}
\label{fig:sdss_quilt_zoom}
\end{figure*}

\begin{table*}
  \centering
  \footnotesize
  \caption{A summary of the characteristics of the volume limited
    samples considered in this paper. 
  Col.~1 - volume label ;
  col.~2 - absolute magnitude range ;
  col.~3 and 4 - minimum and maximum redshift limits respectively;
  col.~5 - sample volume;
  col.~6 - number of galaxies with redshift.
  %col.~7 - number of Jackknife zones considered. 
  Note that \Mstar\ is defined as in Blanton \etal\ (2003b), 
  i.e.\ \Mstar$-5 \log_{10} h = -20.44$. The first three entries in the table 
  correspond to approximately the same volume. 
  }
  \label{tab:samples}
  \begin{tabular}{cccccc} 
    \hline \hline
    Label & M-range & $z_{min}$ & $z_{max}$ & V/$\Mpccube$ & N$_g$ \\ %& N$_{\rm sub}$ \\
    \hline
    {\tt ref} & \Mstar$+0.5\le$M$\le$\Mstar$-0.5$  & 0.0508 & 0.1065 & (258.3)$^3$ & 98 317  \\ % & 16, 25, 49, 100, 225 \\
    {\tt faint-ref} & \Mstar$+0.5\le$M$\le$\Mstar  & 0.0405 & 0.1065 & (263.5)$^3$ & 63 916  \\ % & 16, 25, 49, 100, 225 \\
    {\tt bright-ref} & \Mstar$\le$M$\le$\Mstar$-0.5$ & 0.0508 & 0.1065 & (258.3)$^3$ & 37 914  \\ % & 25, 100, 225 \\
    {\tt bright-all} & \Mstar$\le$M$\le$\Mstar$-0.5$ & 0.0508 & 0.1325 & (325.9)$^3$ & 76 332  \\ % & 16, 25, 49, 100, 225 \\
    {\tt bright-far} & \Mstar$\le$M$\le$\Mstar$-0.5$ & 0.1065 & 0.1325 & (259.0)$^3$ & 38 418  \\ % & 25, 100, 225 \\
    \hline \hline
  \end{tabular}
\end{table*}

\begin{figure*}
\plottwo{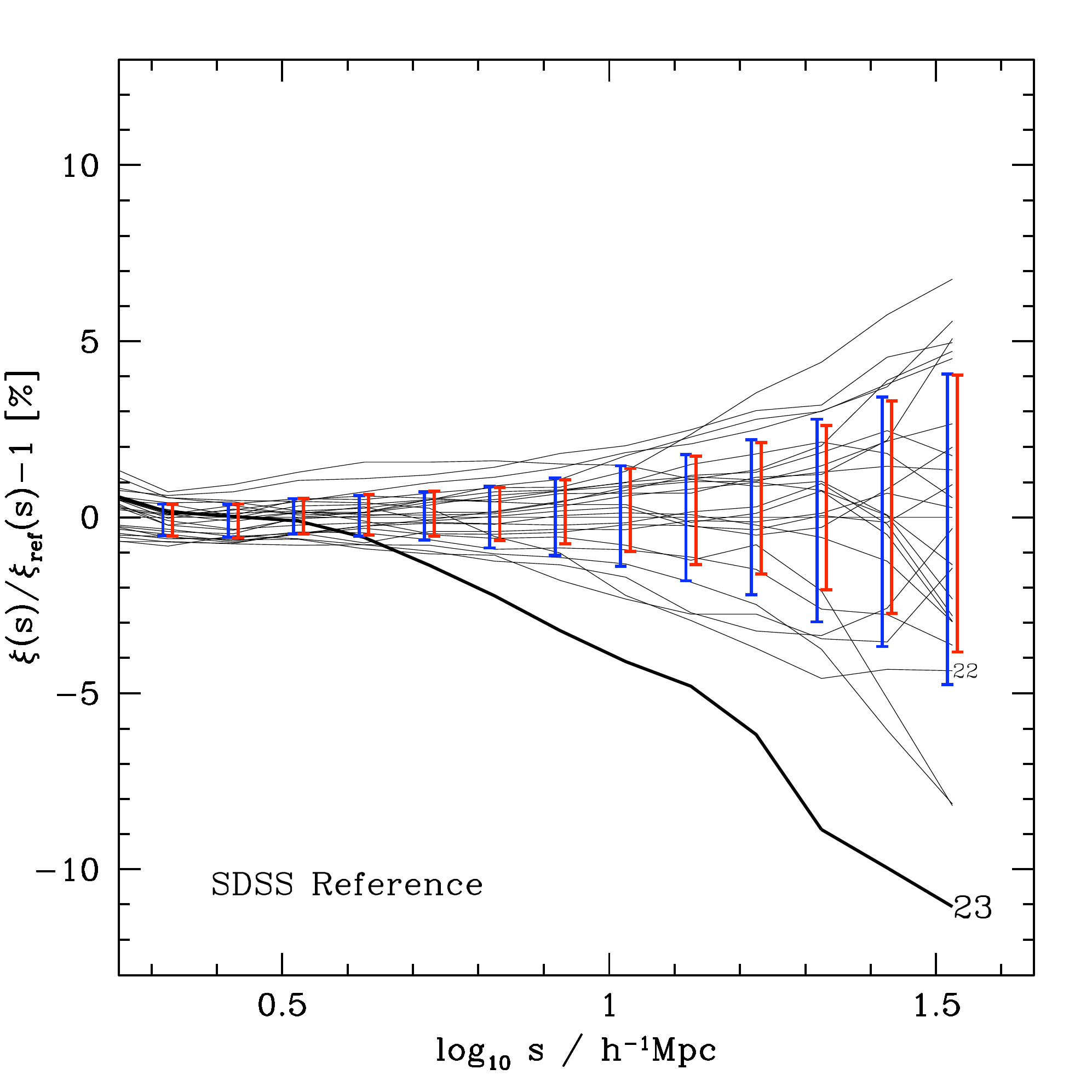}{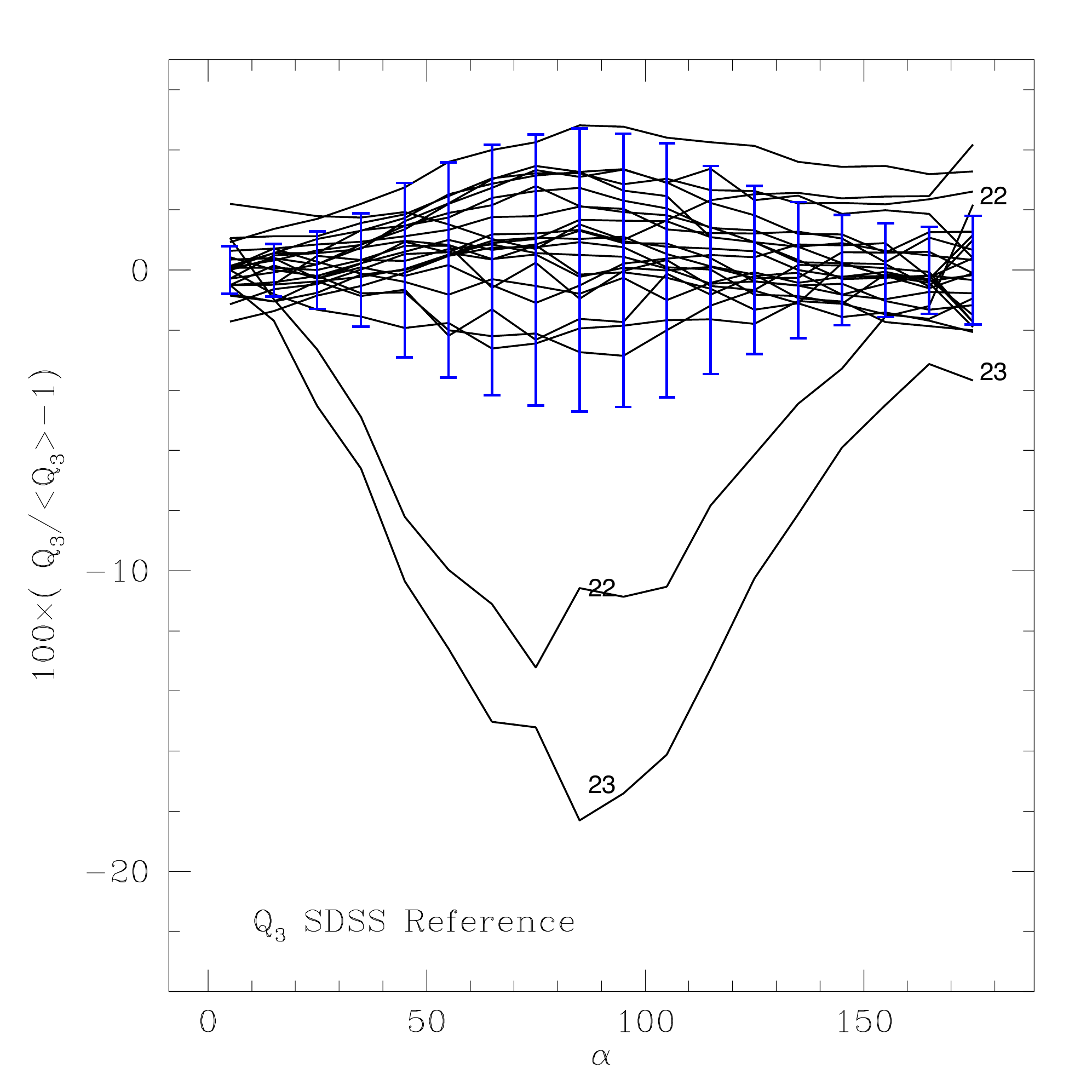}
\caption
{
The distribution of clustering measurements from the Jackknife resamplings 
of the {\tt ref} sample. The left-hand panel shows the measurements for 
$\xi(s)$ and the right-hand panel shows Q$_3(\alpha)$, with $\alpha$
in degrees. In both cases, we plot the ratio of the correlation 
function measured from each Jackknife resampling to that measured 
from the full {\tt ref} sample. 
The lowest amplitude correlation functions are labelled by the zone 
number omitted in the resampling. The larger (blue) errorbars show 
the {\it rms} error estimated from a Jackknife resampling of all of 
the zones, which corresponds to the central 68\% of the distribution 
of measurements. In the left hand panel we also show the slightly smaller
(red) errorbars, corresponding to the {\it rms} error estimated on
omitting zone 23 from the Jackknife resampling altogether.
Note that these errorbars are smaller by a factor 
$\sqrt{{\rm N_{\rm sub}}}$ than those assigned by Jackknife errors to
correlation function measurements (see e.g.\ Eq.~\ref{eq:cov_jk}). 
}
\label{fig:ratio_stats}
\end{figure*}

\subsection{SDSS Jackknife quilts}
\label{sec:jackknife_quilt}

Our goal in this paper is to assess the impact of 
unusual structures on clustering statistics. We do this 
by dividing the SDSS survey up into different regions 
to assess their contribution to the clustering signal. 
A framework in which to do this analysis is provided by the 
Jackknife approach to error estimation (see e.g.\ Norberg \etal\ 2009).  
In this method, a dataset is divided into $N_{\rm sub}$ zones. 
The clustering is then measured from a sample made up of 
$N_{\rm sub}-1$ of these zones, leaving one of the zones out. 
This process is repeated $N_{\rm sub}$ times, leaving out each one of
the zones in turn. The 
scatter between the clustering measurements from the Jackknife samples
is used to estimate the error on the clustering statistic. 
For completeness we quote here the standard relation used to
estimate the Jackknife covariance matrix (see e.g.\ 
Norberg \etal\ 2009, Zehavi \etal\ 2002) when split into N$_{\rm sub}$
samples:  
\begin{eqnarray}
C_{\rm jk}(x_i,x_j) & = & \frac{({\rm N_{\rm sub}}-1)}{{\rm N_{\rm sub}}} \sum_{k=1}^{{\rm N_{\rm sub}}} 
(x_i^k - \bar{x_i}) (x_j^k - \bar{x_j})~,
\label{eq:cov_jk}
\end{eqnarray}
where $x_i$ is the $i^{\rm th}$ measure of the statistic of
interest. It is assumed that the mean expectation value is given by  
\begin{eqnarray}
\bar{x_i} & = & \sum_{k=1}^{\rm N_{\rm sub}} x_i^k / {\rm N_{\rm sub}} \, ,
\label{eq:mean_xi}
\end{eqnarray}
but it can also be given simply by the measurement over the whole
sample. 
Note the factor of ${\rm N_{\rm sub}}-1$ which appears in Eq.~\ref{eq:cov_jk}  
(Tukey 1958; Miller 1974). Qualitatively, this factor takes into account 
the lack of independence between the N$_{\rm sub}$ copies or
resamplings of the data; recall that from one copy to the next, only
two sub-volumes are different (or equivalently ${\rm N_{\rm sub}}-2$
sub-volumes are the same).

Here we divide the SDSS survey into zones in right ascension  
and declination, producing what we call the survey ``quilt''. The
zones have equal areas after taking into account their spectroscopic
completeness, using the completeness mask described above. We use
square 5x5 (i.e.\ $N_{\rm sub}=25$ Jackknife zones) and 15x15 
(i.e.\ $N_{\rm sub}=225$ Jackknife zones) Jackknife quilt patterns in
the right ascension ($\alpha$) and declination ($\delta$) plane,
leading to the quilt patchworks shown in Fig.~\ref{fig:sdss_quilt}. 
We also investigated Jackknife quilts with other dimensions 
(N$_{\rm sub}$= 16, 49, and 100); our results are robust to the 
number of Jackknife zones considered. 

%area= pix * (180/PI)^2 * 4./NP_trans**2 ; NP_trans:3000
%Nsub=25:
%<npix>: 191711.7969 2242.507812 0.01169728674    -> 279.7+/-3.3
%<rpix>: 171340.0625 0.4777350724 2.788227448e-06 -> 249.97 +/- 
%Nsub=225:
%<npix>: 21301.31055 427.4923401 0.02006882802    ->  31.1+/-0.6
%<rpix>: 19037.78516 0.3615865111 1.899309609e-05 ->  27.7

For the SDSS footprint considered in this paper, $\sim6250$~deg$^2$, 
each zone in the 25 Jackknife quilt covers a completeness weighted
area of precisely $250$~deg$^2$, while the average area subtended by a
zone is $279.7\pm3.3$~deg$^2$. The corresponding numbers for the 225
Jackknife quilt are $27.7$~deg$^2$ and $31.1\pm0.6$~deg$^2$. Because
of the high spectroscopic completeness threshold considered here and
the uniform tiling of the SDSS survey the relative {\it rms}
variance on the area subtended by each Jackknife zone is less than 2
per cent.

\subsection{Volume limited samples}

In this paper we focus our attention on galaxies around \Mstar,  
which in the SDSS $r$-band corresponds to 
\Mstar$-5 \log_{10} h = -20.44$ (Blanton \etal\ 2003b). In order 
to obtain an absolute magnitude for each galaxy, and also to 
construct volume limited samples, we need to adopt 
a $k+e$-correction. We apply a global $k+e$-correction 
to a nominal reference redshift $z_{\rm ref}=0.1$ following 
Blanton \etal\ (2003a,b). A volume limited sample is defined 
by the apparent magnitude range of the survey and a bin in 
luminosity: a galaxy in a given luminosity bin would fall within the
apparent magnitude range of the survey if placed at any redshift
between the limits $z_{\rm min}$ and $z_{\rm max}$.  
Here we consider a variety of volumes and luminosity bins close to 
\Mstar, and their basic properties are listed in
Table~\ref{tab:samples}. A brief description of each sample is as
follows:  
\begin{itemize}
\item {\tt ref} - the volume limited sample for galaxies in a one
  magnitude wide bin centred on \Mstar.
\item {\tt bright-ref} - the bright half magnitude bin of the 
  {\tt ref} sample.
% galaxies in the half-magnitude wide bin including 
% \Mstar\ and brighter, covering the same volume as the {\tt ref} sample.
\item {\tt faint-ref} - the volume limited sample for galaxies in the 
   half-magnitude bin fainter than \Mstar: this volume is $\sim 6$ per
   cent larger than the one of the {\tt ref} sample.
\item {\tt bright-all} - the volume limited sample for galaxies
  in the half-magnitude bin brighter than \Mstar. By construction this
  sample extends to a higher redshift than the {\tt bright-ref} sample. 
\item {\tt bright-far} - the high redshift half of the 
  {\tt bright-all} volume. The {\tt bright-ref} and {\tt bright-far}
  samples combined give the {\tt bright-all} sample. 
\end{itemize}
In summary, the {\tt ref} and {\tt bright-ref} volumes are
identical; the {\tt ref} and {\tt faint-ref} volumes differ by only
$\sim 6$ per cent; the {\tt ref} and {\tt bright-far} volumes are
fully disjoint but cover (to within less than one per cent) an
equal volume of space, i.e.\ $\sim 0.017 \Gpccube$. 
The CasJobs SQL query used to generate the input catalogue from which
these volume limited samples are constructed is given in
Appendix~\ref{sec:appendix_query}. 

In Fig.~\ref{fig:sdss_quilt_zoom} we show the full (top) and
zoomed in view (bottom) of the
angular galaxy distribution of the {\tt ref} volume limited sample
viewed through the completeness mask, together with the boundaries of
the zones of the 5x5 and 15x15 Jackknife quilts (thick red and thin
blue lines respectively). Big empty patches, like the one at $\alpha
\simeq 175$deg.\ and $\delta \simeq 6$deg., corresponds to areas masked
by the survey completeness mask. This plot shows clear evidence of
large coherent galaxy structures, spanning several zones: this is
precisely the type of large scale structures whose influence on
clustering statistics we aim to investigate in the paper.

\begin{figure}
\plotone{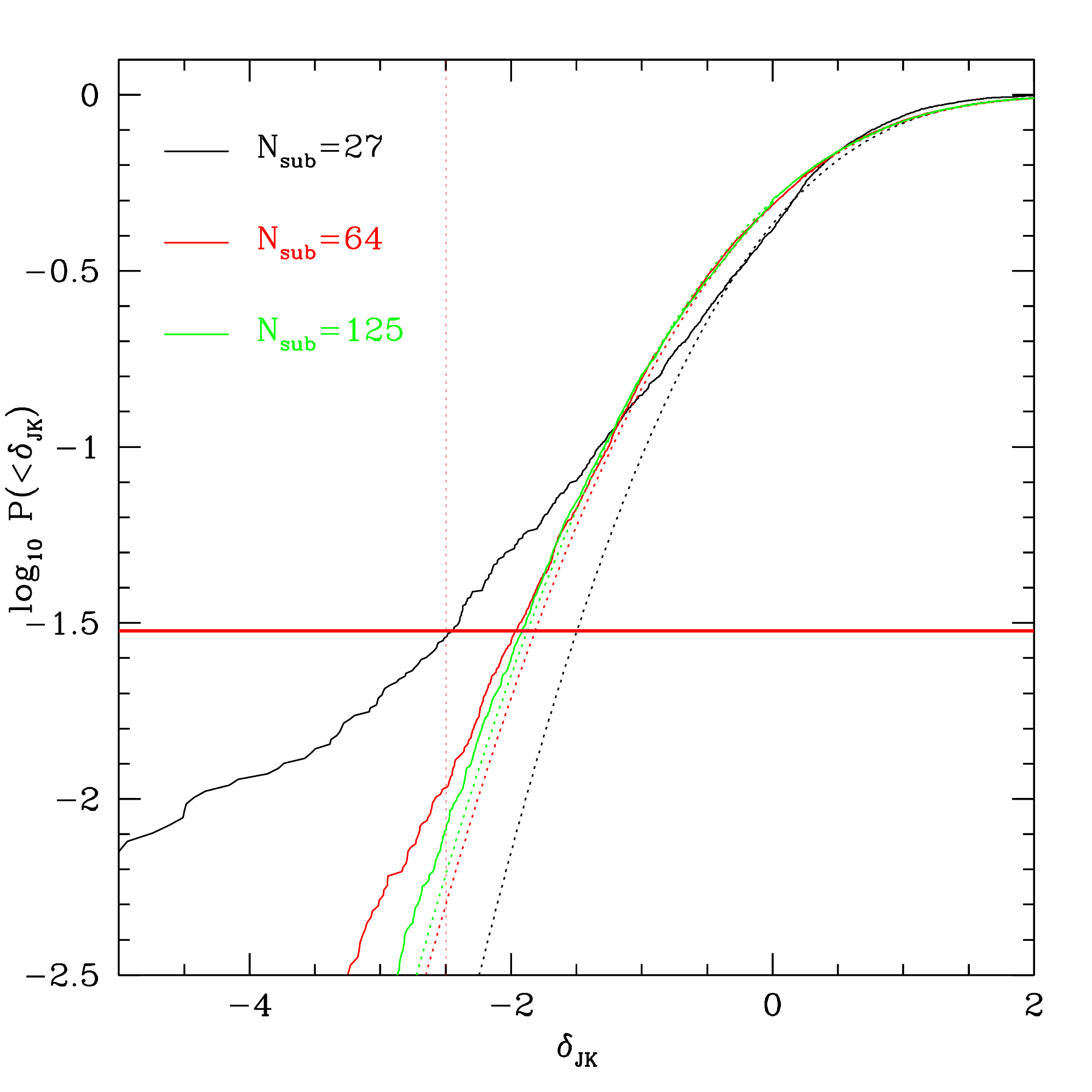}
\caption
{
The probability of finding a JK ensemble fluctuation, 
$\delta_{\rm JK}$, below some value 
for the distribution of clustering measurements from 
an ensemble of SDSS mocks made from N-body simulations (see 
Section~\ref{sec:lcdm_sim}) 
for three values of \nsub (solid lines). 
The $\delta_{\rm JK}$ distribution is estimated here for $\xi(s)$ with
\nsub=27, 64 and 125 (black, red and green respectively) over the range
$12-14\Mpc$.  
The corresponding Gaussian distributions (same colours, but dotted
lines) have the same median value and a variance which encloses 68\%
of the simulation measurements.
The JK ensemble fluctuation for small
number of JK zones (\nsub=27) present a non negligible non-Gaussian
tail, highlighting the 
importance of using 
N-body simulations to estimate its cumulative probability
distribution function, while for larger \nsub\ values the distribution
can be well approximated by a Gaussian.
}
\label{fig:JK_new}
\end{figure}

\subsection{Clustering estimators}

In this paper we consider the spherically averaged 2-pt correlation 
function, $\xi(s)$, and the reduced 3-point correlation function,
Q$_3(\alpha)$.  The clustering estimators used are described in detail
in Norberg \etal\ (2009) and Gazta\~naga \etal\ (2005), for $\xi(s)$
and Q$_3(\alpha)$ respectively.  

In the case of $\xi(s)$, we count all pairs out to a maximum
separation in redshift space of $60 \Mpc$. For Q$_3(\alpha)$ we
consider triplets in which one side of the triangle is $8\Mpc$ and the 
other is $16\Mpc$, with an opening angle of $\alpha$ between 
these two sides. This is one of the many configurations 
considered in Gazta\~naga \etal\ (2005) (see their Fig.~1 for an 
illustration of the triplet). We have focused on a minimum of $8\Mpc$
scales for SDSS (as opposed to $6\Mpc$ in 2dFGRS) because the larger
SDSS volume makes it possible to explore larger (weakly non-linear) scales.
We follow the implementation of the 
Jackknife method outlined in Norberg \etal\ (2009), and briefly
summarized in subsection~\ref{sec:jackknife_quilt}.

We estimate $\xi(s)$ and Q$_3(\alpha)$ for all of the volume 
limited samples listed in Table~\ref{tab:samples}, as well as 
for the Jackknife resamplings of the different dimension quilts
considered in this paper, i.e.\ mainly N$_{\rm sub}$= 25 and 225, but
also N$_{\rm sub}$= 16, 49, and 100.

\begin{figure*}
\plottwo{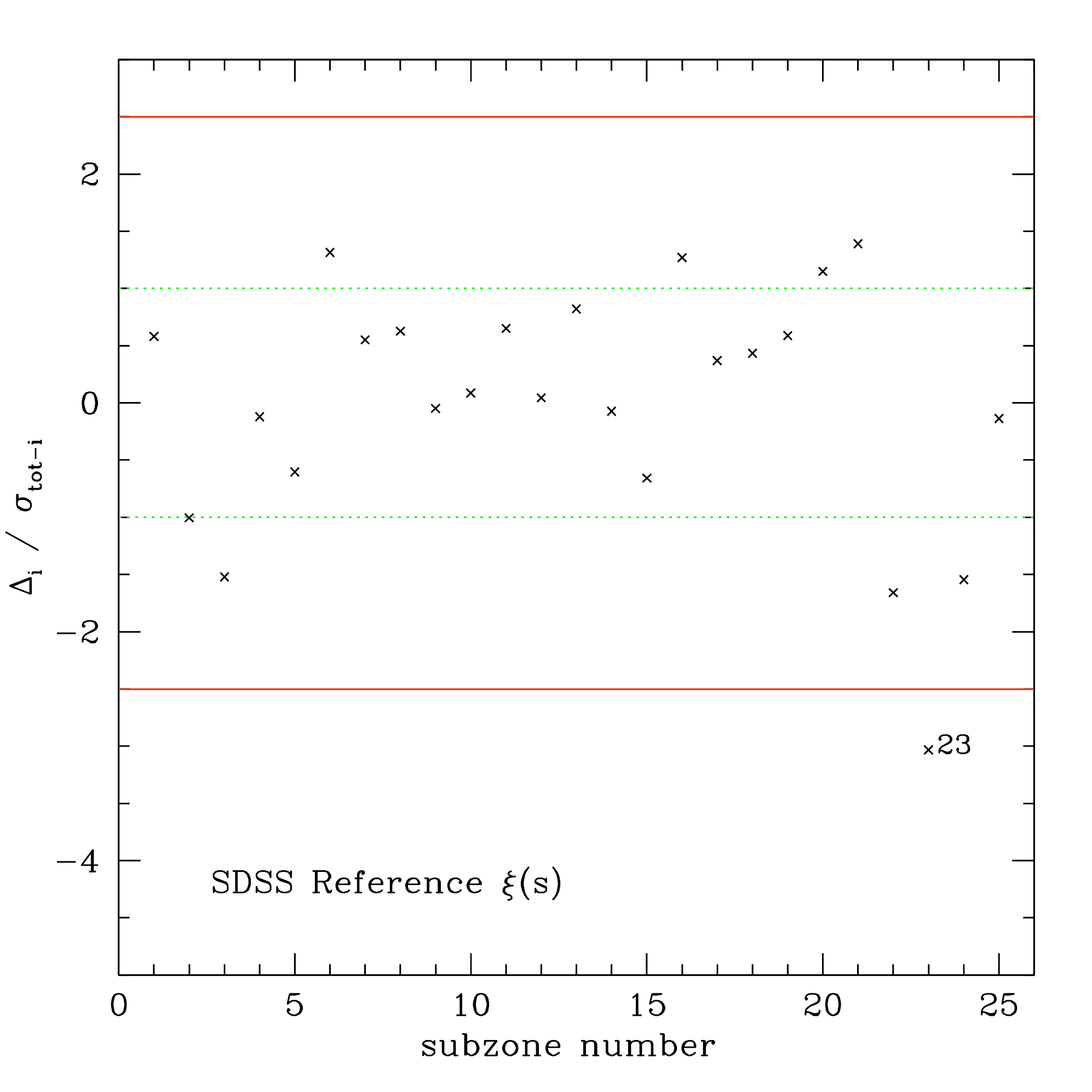}{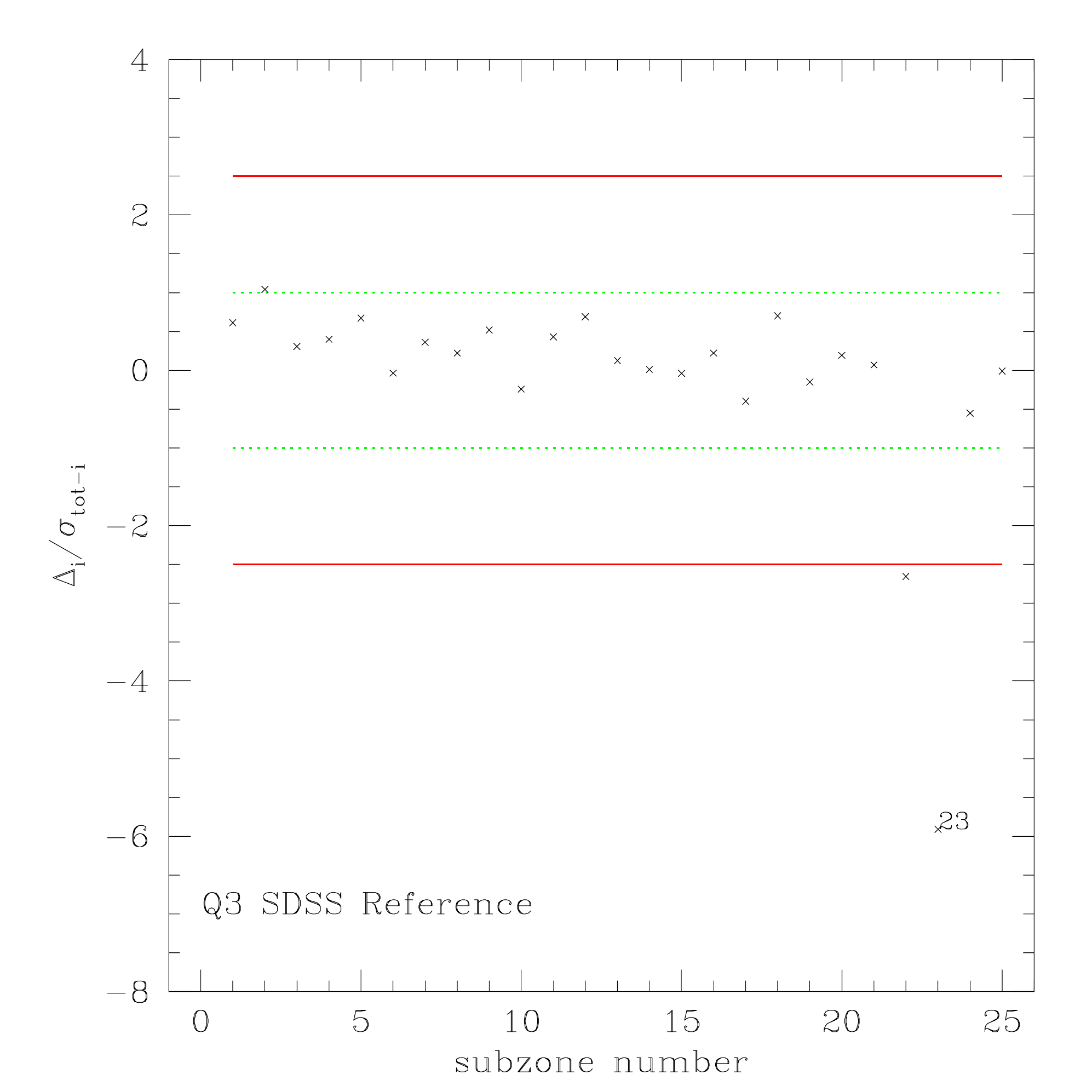}
\caption
{
The JK ensemble fluctuation, $\delta_{JK}^{i}$
(as defined in Eq.~\ref{eq:JK_fluctuation}) using pair separations of
$12-14\Mpc$ in $\xi(s)$ (left panel) and triplets separated by similar
scales for 
Q$_3(\alpha)$ (right panel). Each point corresponds to leaving out 
the zone number plotted on the x-axis. This new statistic is shown in
both panels for a \nsub=25 Jackknife quilt, as shown on the left panel
of Fig.~\ref{fig:sdss_quilt}. 
The fluctuation is plotted using a cross if its value lies within 
$\pm 3$; otherwise the point is labelled with the zone number that 
was omitted.
}
\label{fig:JK_fluctuation}
\end{figure*}

\section{Identification of outlying regions}
\label{sec:jk_outliers}

We use a Jackknife technique to identify survey regions 
which have a big influence on the measured clustering signal. 
The method consists of comparing clustering 
measurements from each of the Jackknife resamplings 
of the data with that obtained from the 
full dataset. If the measurement from a particular 
resampling is unusual next to the measurements obtained 
from the others, then the omitted zone is 
referred to as an outlying zone. 
The key here is the definition of ``unusual''. 
The very presence of an outlying zone 
affects the distribution of measurements from the Jackknife 
resamplings, with the consequence that simple concepts which 
apply to Gaussian distributions, such as the mean and variance, 
need to be redefined. To deal with this we  
introduce two new statistics to quantify outliers: the JK 
resampling fluctuation and the JK ensemble fluctuation.

In this section we apply the Jackknife resampling to the {\tt ref} 
sample listed in Table~\ref{tab:samples}. We consider the clustering
in the other samples listed in Table~\ref{tab:samples} in
Section~\ref{sec:galform_impact}.

\subsection{The JK resampling fluctuation}

In Fig.~\ref{fig:ratio_stats} we plot the two and three 
point correlation functions measured from each of the  
Jackknife resamplings of the {\tt ref} sample. We show 
the distribution of these measurements by plotting 
the scaled quantity $\Delta_{i}$ which is defined as 
\begin{equation}
\Delta_{i} = \frac{x_{i}}{x_{\rm full}} - 1, 
\label{eq:JK_ratio}
\end{equation}  
where $x_{i}$ represents a clustering statistic, which in the present
paper will be either the two or three point function 
measured from a JK resampling of the data on omitting zone $i$ 
and $x_{\rm full}$ indicates the corresponding measurement using 
the full sample. We call this quantity the {\it JK resampling fluctuation} 
as it quantifies the deviation of the clustering measured on omitting 
a particular zone from the measurement using the full dataset.

In Fig.~\ref{fig:ratio_stats} we use 25 Jackknife zones. 
The left panel of Fig.~\ref{fig:ratio_stats} shows the 
scatter in the two-point correlation function measured from 
the different resamplings and the right hand panel shows the scatter 
in Q$_3(\alpha)$. The measurements of $\xi(s)$ and Q$_3(\alpha)$
when omitting zone 23 (which is indicated in
Fig.~\ref{fig:sdss_quilt_zoom}) are clear outliers compared to the
measurements from the other jackknife resamplings. 
Both of these measurements lie well outside the variance 
of the resamplings which 
is indicated in the plot by the error bars. Note that the 
variance here corresponds to the range which encloses the 
central 68\% of the JK measurements.
Of course, when interpreting this plot it is important to 
remember that bins of pair separation or 
opening angle are correlated. Furthermore the measurements 
from different resamplings are also correlated, as they  
differ by only 8\% in area and hence by the same amount in 
volume when using 25 Jackknife zones.  

\subsection{The JK ensemble fluctuation}

The fluctuation in the clustering signal on omitting zone 23 
is in fact even more significant than suggested by 
Fig.~\ref{fig:ratio_stats}. 
To demonstrate this we now apply the Jackknife technique to the 
data in a slightly different way from the standard approach 
as outlined in Norberg et~al (2009). 
The {\it relative} {\it rms} error estimated in the standard 
Jackknife scheme by splitting the data into \nsub\ zones  
is denoted by $\sigma_{\rm tot}$:
\begin{equation}
\sigma_{\rm tot}^2= {1\over{N_{\rm sub}}} \sum_{i=1}^{N_{\rm sub}} \Delta_i^2 ,
\label{eq:sigma_tot}
\end{equation}
where $\Delta_i$ is the JK fluctuation defined by Eq.~\ref{eq:JK_ratio}.
\footnote{Note that we do not multiply here by $N_{\rm sub}-1$, as would 
be necessary to scale this variance to the one corresponding to the full
sample (Norberg \etal\ 2009). Here we are interested instead in 
how the Jackknife errors fluctuate from one resampling to another.}
We then make a similar error estimate from a new dataset, 
which is comprised of \nsub$- 1$ zones, omitting the $i^{\rm th}$
zone from the set altogether:
\begin{equation}
\sigma_{\rm tot-i}^2= {1\over{N_{\rm sub}-1}} \sum_{j\neq i}^{N_{\rm sub}-1} \Delta_j^2 ,
\label{eq:sigma_i}
\end{equation}
 The Jackknife resampling is done in this instance sampling only from 
the \nsub$ - 1$ zones, without considering the $i^{\rm th}$ zone at all. 
The {\it rms} error obtained in this way is written as $\sigma_{\rm tot-i}$. 

In order to quantify how the scatter in the clustering depends upon
which zone is left out, we introduce a new statistic,
the JK ensemble fluctuation, $\delta_{JK}^{i}$,
defined as:  
\begin{equation}
\delta_{JK}^{i} \equiv {\Delta_i\over{\sigma_{\rm tot-i}}}, 
\label{eq:JK_fluctuation}
\end{equation}
which is just the JK resampling fluctuation for the $i^{\rm th}$ zone
normalized to its corresponding {\it rms} error, i.e.\ 
$\sigma_{\rm tot-i}$. 

The probability of finding a particular value of the JK ensemble fluctuation  
is plotted in Fig.~\ref{fig:JK_new} for a distribution which is very
similar to that expected for the SDSS data, derived from an ensemble
of N-body simulations described in Section~\ref{sec:lcdm_sim}. 
In Fig.~\ref{fig:JK_new} $\delta_{\rm JK}$ is estimated for $\xi(s)$
over $12-14\Mpc$, the range of scales most relevant for this paper. 
For comparison purposes, we plot a
Gaussian that has the same median value as that obtained from the
simulations and has a variance which corresponds to the range which
encloses 68\% of the correlation functions measured from the simulations.
These distributions describe the expected range of fluctuations in the
new JK ensemble fluctuation. For small values of 
$\delta_{JK}^{i}$ the two distributions are similar, due to the way the 
Gaussian has been set up. However, they differ appreciably in the tails. 
In the case of the Gaussian, a value of $\delta_{JK}^{i}< -1.5$ would 
be obtained 3\% of the time, whereas for the distribution from 
the simulations, we would expect to see such a value in 
around 8\% of cases; the 3\% chance in the simulation case corresponds 
to $\delta_{JK}^{i} < -2.5$. The non-Gaussian nature of the
distribution of the JK ensemble fluctuation decreases with the number
of JK samples considered, and approaches asymptotically the
corresponding Gaussian distribution for $\delta_{JK}$.
Therefore for small (large) number of JK samples, one has to
consider the non-Gaussian (Gaussian) distribution 
to estimate the significance level of a given $\delta_{JK}^{i}$.

In Fig.~\ref{fig:JK_fluctuation} we use the new statistic, $\delta_{JK}^{i}$, 
defined by Eq.~\ref{eq:JK_fluctuation}, to quantify the significance of 
the fluctuations in the clustering measurements. The error is calculated 
using pair separations in the range $12-14\Mpc$.\footnote{Any scale could 
be considered for this statistic, but we decide here to focus on one range 
appropriate for both statistics; with Q$_{3}(\alpha)$ sampling scales 
between $8\Mpc$ and $24\Mpc$, it is natural to settle on the range 
$12-14\Mpc$.}
In the left hand panel we plot JK ensemble fluctuation for $\xi(s)$ on
splitting the data into 25 Jackknife zones and in the right we show 
the corresponding result for Q$_3(\alpha)$, sampling similar scales.
Fig.~\ref{fig:JK_fluctuation} (left-hand panel) shows that if the
Jackknife zones were truly independent, then the fluctuation in the
{\it rms} error on leaving out zone 23 corresponds to $\delta_{JK}^{i}<-3$, 
which, according to Fig.~\ref{fig:JK_new}, should occur around 1.6\% 
of the time. For a Gaussian distribution, this large value of
$\delta_{JK}^{i}$ is even less likely. 
We note that this probability comes from the fraction of JK
regions from an ensemble of simulations that have $\delta_{\rm JK}<-3$, as
$\delta_{\rm JK}$ is estimated for each JK region. In other
words, the constraint is on the probability of a JK region being this
(or more) extreme and not on the probability that a survey contains
such an extreme region.  
From the simulations we infer that $\sim50$ percent of SDSS {\tt ref}
like volumes would contain one or more JK regions with  
$\delta_{\rm JK}<-3$.
For Q$_3(\alpha)$ (right-hand panel of Fig.~\ref{fig:JK_fluctuation}) the fluctuation in the {\it rms} error on leaving out zone 23
corresponds to $\delta_{\rm JK}^{i} \sim -6$, which translates into a
probability of less than $\sim0.4$\% (see
Fig.~\ref{fig:JK_new}), 
assuming the distribution of the JK ensemble fluctuation is the same
for $\xi(s)$ and Q$_3(\alpha)$. It should be noted that the
$\delta_{\rm JK}$ distribution should be dependent on the statistic
considered, as well as on the number of zones considered. In our
comparisons we always find that outliers in Q$_3(\alpha)$ tend to
correspond to the ones in $\xi(s)$. Hence we assume here that the
distribution of $\delta_{\rm JK}$ is similar for the two and three
point correlation functions.

For both the two and three point correlation functions, 
the measurements on omitting zone 23 are 
significant outliers, 
not only visually as in Fig.~\ref{fig:ratio_stats} but also
statistically when interpreting their $\delta_{JK}$ value (less than
$-3$ for both $\xi(s)$ and Q$_3(\alpha)$) in terms of the
global $\delta_{JK}$ distribution as extracted from the simulations.
It is not surprising to see that the significance 
level is much larger for the higher order statistic 
(i.e.\ Q$_3(\alpha)$), 
as it is well known 
that such statistics are much more sensitive to large scale fluctuations 
than the two point function.  We note that zone 22 is also an outlier for the
reduced 3-point statistic, albeit to a much lesser extent than zone
23. This is only really noticeable with this new statistic, as in the
right hand panel of Fig.~\ref{fig:ratio_stats} the first impression
one has is that zones 22 and 23 are both outliers to roughly a similar
extent.

\begin{figure}
\plotone{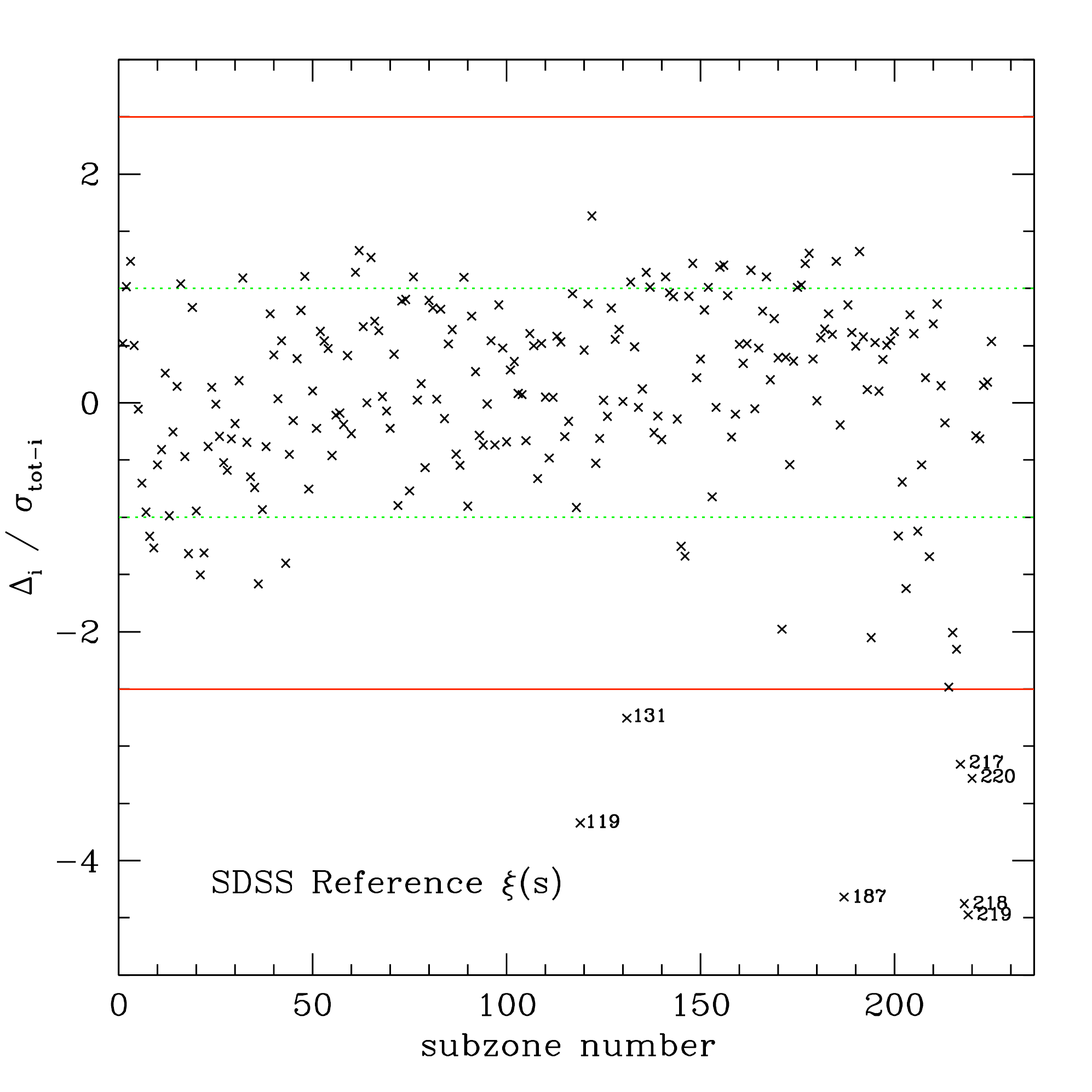}
\caption
{
The same plot as in the left panel of Fig.~\ref{fig:JK_fluctuation},
but for \nsub=225, i.e.\ for the 15x15 Jackknife quilt, as shown in the
right panel of Fig.~\ref{fig:sdss_quilt}. For clarity, crosses have
been omitted whenever $|\delta_{\rm JK}|>2.5$. There are 
7
zones which present significant deviations from the expectation in the
JK ensemble fluctuation statistic (as defined in
Eq.~\ref{eq:JK_fluctuation}) using pair separations of
$12-14\Mpc$.
For this large \nsub, the $|\delta_{\rm JK}|>2.5$
threshold corresponds to a 2.5-$\sigma$ level, implying that the
Gaussian expectation for the number of outliers is just
$\sim1.4\pm1.2$, significantly lower than the number of observed
outliers.    
}
\label{fig:JK_fluctuation_225}
\end{figure}

Fig.~\ref{fig:JK_fluctuation_225} presents the JK ensemble fluctuation
for the \nsub=225 quilt. Zones 187 and 217 
to 220 stand out as much as zone 23 did for the \nsub=25
quilt. These zones from the 225 Jackknife quilt are mostly co-spatial
and overlap mainly with zones 22 and 23 (and a small amount with zone 24)
from the 25 Jackknife quilt, as shown more explicitly in
Fig.~\ref{fig:sdss_quilt_zoom}. 
Together they map out the central parts of the SDSS great wall 
(Gott \etal\ 2005), a feature uncovered partially in the earlier
2dFGRS analysis of Baugh \etal\ (2004). Out of the seven outliers in
the 225 Jackknife quilt, five belong to this structure.

\subsection{Blind search and number of JK}

The variant on the Jackknife technique we have described in this
section is an objective way to find unusual structures in the galaxy
distribution. The Jackknife zones provide a blind test  
to detect superstructures based on uncovering outliers in 
clustering measurements,
although it is unlikely that this method would be the preferred one
for actually detecting superstructures. It is certainly more valuable
as test of the homogeneity of the JK subsampling chosen.
Furthermore, we have introduced a statistic which allows us to 
quantify the significance of the outliers. Another feature of 
this approach is that it can be used to suggest the appropriate 
size of Jackknife zone to use in error estimation. If, for example, we 
find that several adjacent zones give outlying clustering results, 
as is the case for the 225 Jackknife quilt above, then  
the analysis should be repeated with larger zones. 
In this way, either the outliers 
disappear or the unusual clustering is restricted to one zone, as 
we found for the 25 quilt Jackknife. With one outlier zone, one could then 
present the clustering analysis both with and without this zone to 
show its impact on the results (as was done in our earlier
higher-order clustering analyses of the 2dFGRS, e.g.\ Baugh
\etal\ 2004; Croton \etal\ 2004, 2007; Gazta\~naga \etal\ 2005).
% see also Section~\ref{sec:SDSS_2dFGRS}).
%
Similarly, Zehavi et al. (2005, 2011) presented SDSS results for a
full \Mstar\ galaxy sample and for a sample with a redshift limit
designed to exclude the Sloan Great Wall.

\subsection{Impact on errors}

Another consequence of the analysis in this section is the 
impact of zones containing unusual structures on the error estimated 
using the Jackknife method. Fig.~\ref{fig:ratio_stats} shows that the
Jackknife resampling which produces an outlying cluster statistic  
can lead to a over estimate of the variance. The blue errorbars 
in Fig.~\ref{fig:ratio_stats} show the variance estimated using 
all of the Jackknife resamplings, including the outlier. The red errorbars 
show the variance estimated from a Jackknife resampling of 
24 zones in the 25 patch quilt, leaving out zone 23 altogether. 
For the two point function, there is a modest reduction in the 
variance on leaving out the zone containing the superstructure, but
more importantly there is a significant uncertainty in the
corresponding Jackknife covariance matrix when one sample is the
source of most of the variance. This becomes clear in
Fig.~\ref{fig:JK_fluctuation} for both $\xi(s)$ and Q$_3(\alpha)$,
where one notices a systematic bias in the scatter in the JK ensemble
fluctuation.
If all samples were equally important for the variance, this new
statistic would be distributed symmetrically around the expectation
value. This is not the case for either of the clustering statistics 
considered here.

\begin{figure*}
\plottwo{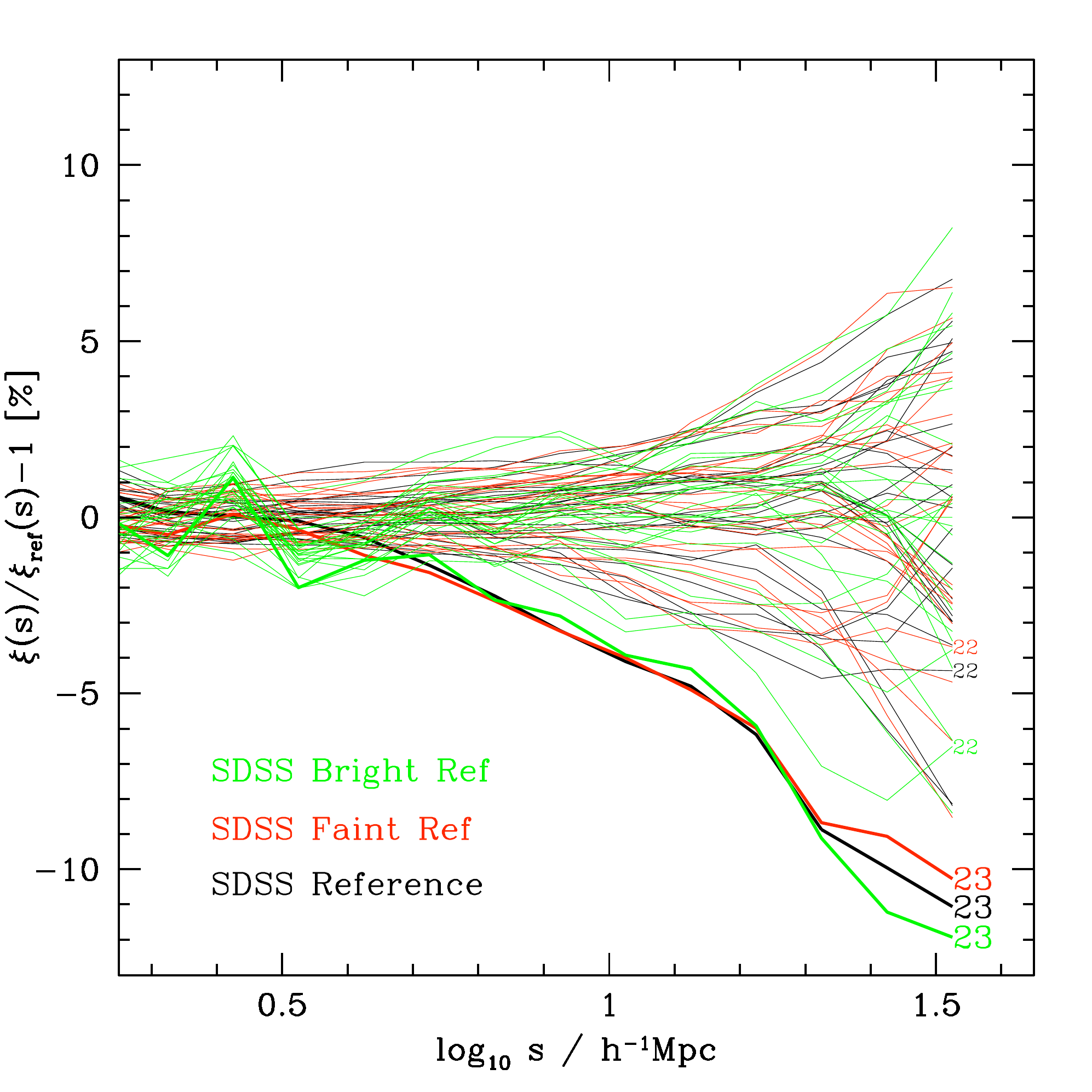}{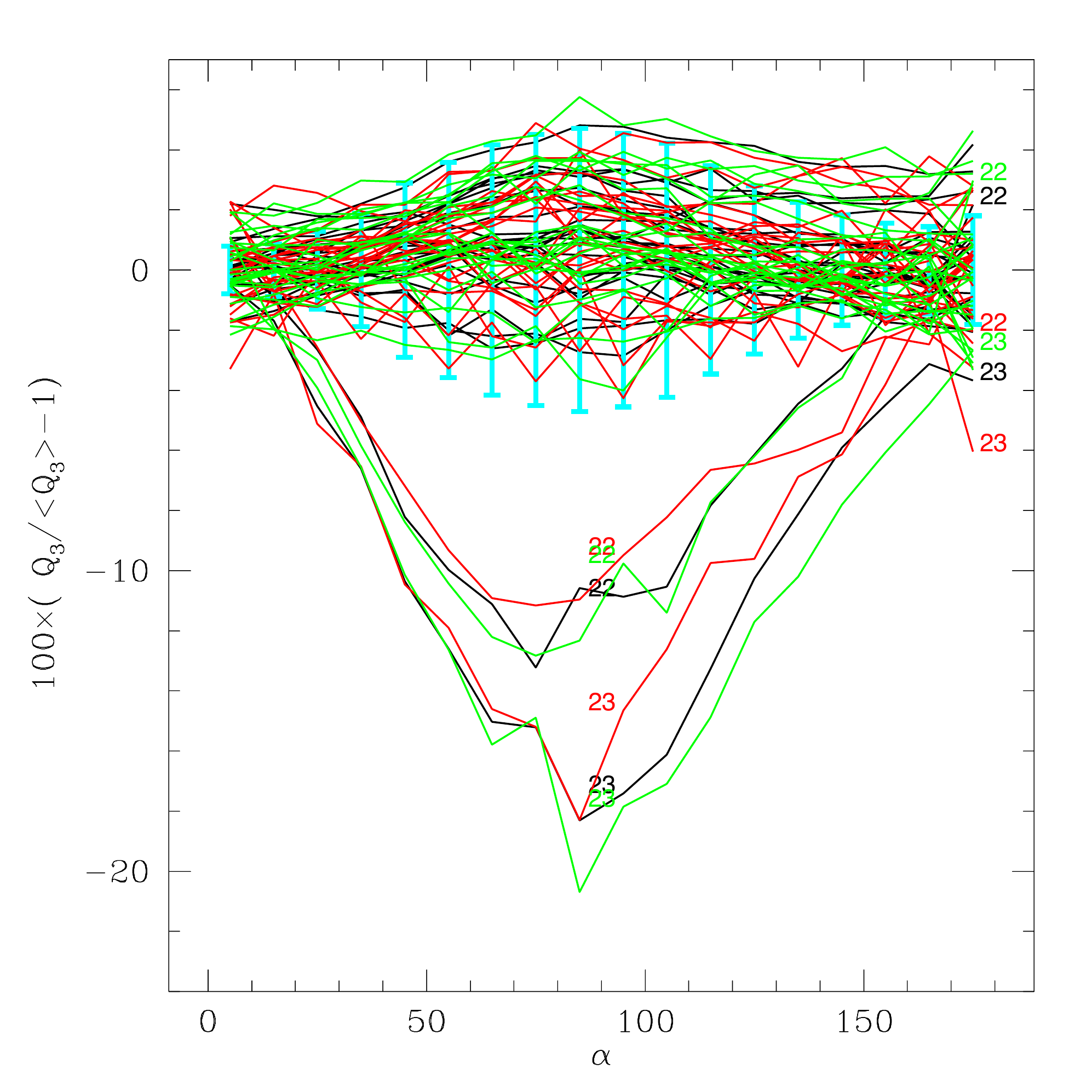}
\caption
{ 
Clustering statistics measured using Jackknife resamplings of  
galaxies from the same volume, using the {\tt ref} sample (black), 
the {\tt bright-ref} (green) and {\tt faint-ref} (red). The left-hand 
panel shows $\xi(s)$ and the right-hand panel Q$_3(\alpha)$. 
The most significantly outlying clustering statistics are labelled by
the omitted zone number (for clarity we place labels both on the
center and on the right of each line for $Q_3$). The errorbars
correspond to the {\it rms} clustering error from the Jackknife
resamplings of the {\tt ref} sample. 
}
\label{fig:galform_impact}
\end{figure*}

%\section{Understanding the influence of superstructures} 
\section{The influence of superstructures} 
\label{sec:galform_impact}

Once we have identified a superstructure using the Jackknife 
approach set out in the previous section, we can study its 
influence on the clustering measured from other volume limited 
samples drawn from the survey. Croton \etal\ (2005) found that 
the 2dFGRS superstructures had an impact only on the \Mstar\ 
volume limited sample, 
with Zehavi et al. (2005) reaching almost the same conclusion from
their SDSS analysis. 
In the case of the sample one magnitude 
brighter than \Mstar, the clustering measurements did not show 
any unusual features, even though the superstructure was also 
included within the volume. Two things change when moving from 
the \Mstar\ volume limited sample to a sample that is 
one magnitude brighter: the volume of the sample increases and 
the way in which structures are represented changes. Importantly, the
number of galaxies tracing the superstructure is different between 
the two volume limited samples, and so the contrast of the 
superstructure will also be different. Also, a larger volume could 
result in other superstructures being included, diluting the impact 
of any one structure on the measured clustering.  

In an attempt to disentangle and understand these effects 
(namely, the increase in the volume used to measure clustering and the
representation of structures with different numbers of galaxies) 
we compare clustering measurements in a range of volume 
limited samples (which are listed in Table~\ref{tab:samples}). All our
comparisons use relative clustering statistics for a given sample,
allowing us to properly compare different galaxies without the need to
model tracer dependent biases, such as luminosity or colour dependent
clustering.

First, we look at different tracers within the same volume, to isolate
the impact of the sampling of the superstructure or its contrast
relative to the other structures within the volume. In
Fig.~\ref{fig:galform_impact}, we compare the clustering measured from
Jackknife resamplings of the three samples: the {\tt ref} (one
magnitude bin centred on \Mstar), the {\tt bright-ref} (the half
magnitude bin brighter than \Mstar) and {\tt faint-ref} (the half
magnitude bin fainter than \Mstar) samples. These galaxy samples are
constrained to cover approximately the same volume and hence are
subject to the same underlying density fluctuations. The only
difference between them is the number of galaxies which populate the
structures within the volume. Fig.~\ref{fig:galform_impact} shows that
the identity of the outlying zone is not sensitive to which galaxies
trace out the superstructure within the same volume. The left hand
panel of Fig.~\ref{fig:galform_impact} shows that zone number 23 from
the 5x5 Jackknife quilt is a clear outlier for all three samples.
The right hand panel of Fig.~\ref{fig:galform_impact} shows the same 
comparison between samples for Q$_3(\alpha)$. For this statistic, 
the omission of two zones, either number 22 or 23, leads to
measurements which stand out for all three samples. Hence, the
influence of the superstructure is not affected by the number of
galaxies used to trace it within a fixed volume, nor by the tracer (to
within the limitation of the comparison done here, with galaxies
spanning just one magnitude in brightness).

\begin{figure}
\plotone{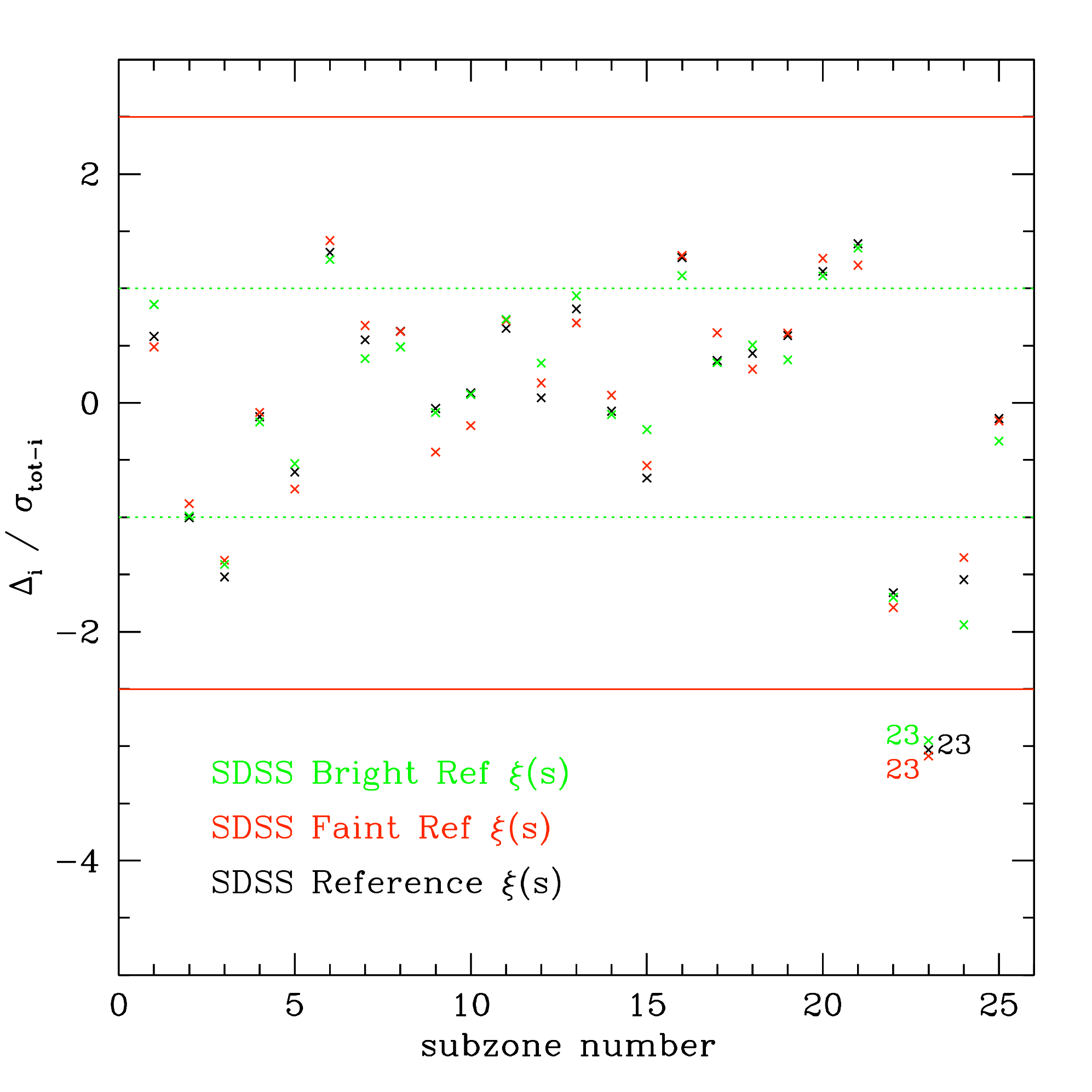}
\caption
{
The JK ensemble fluctuation for $\xi(s)$ measured from 
the {\tt ref} (black), {\tt faint-ref} (red) and {\tt bright-ref} (green) 
volume limited samples for \nsub = 25. 
}
\label{fig:JK_fluctuation_galform}
\end{figure}

To put our conclusions on a quantitative footing, 
we show in Fig.~\ref{fig:JK_fluctuation_galform} 
the corresponding JK ensemble fluctuation of $\xi(s)$ 
for the three samples discussed in Fig.~\ref{fig:galform_impact}. 
As expected, the JK ensemble fluctuation computed for 
$\xi(s)$ is virtually indistinguishable for the three samples 
extracted from the same \Mstar\ volume. Very similar results 
are found for $Q_3$.

\begin{figure*}
\plottwo{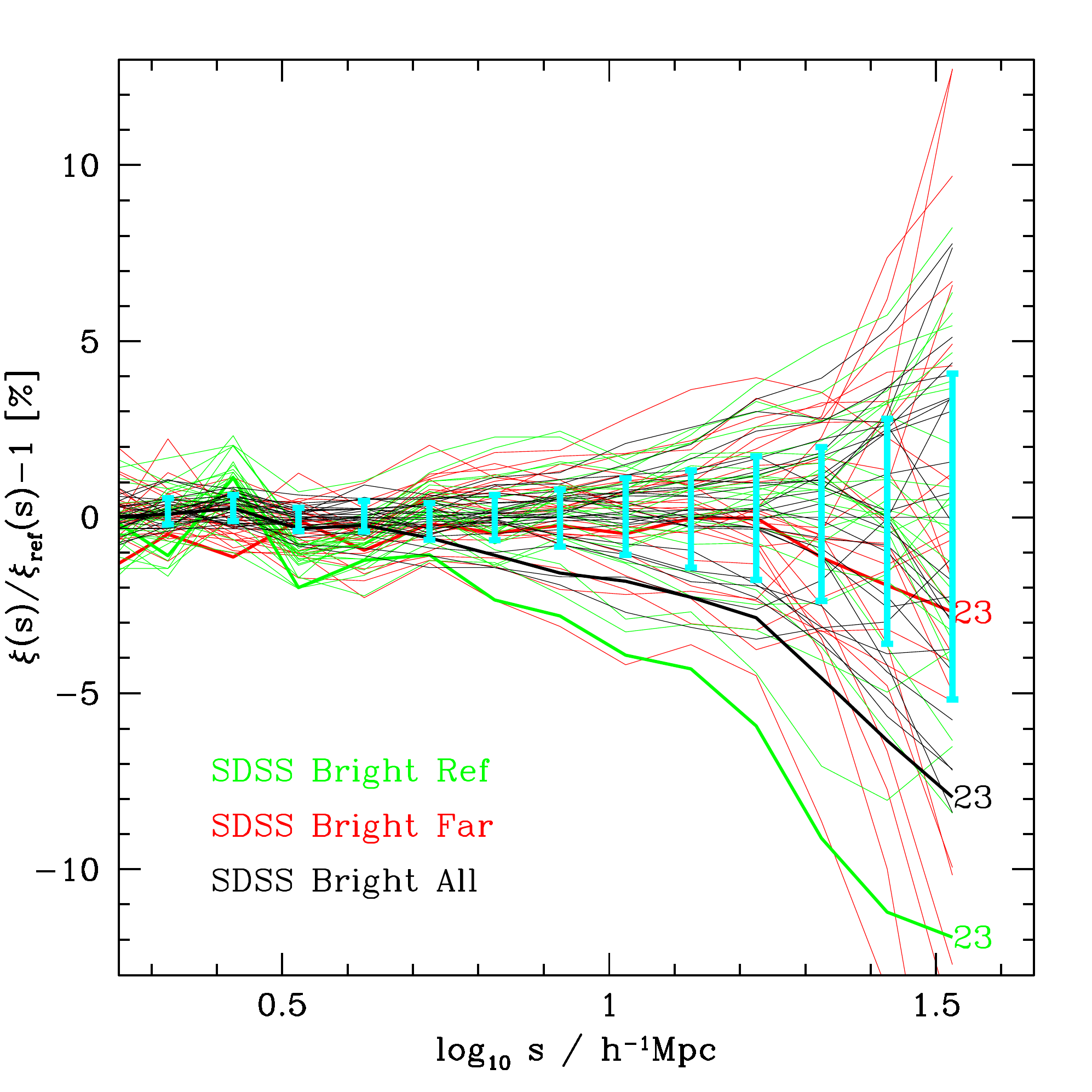}{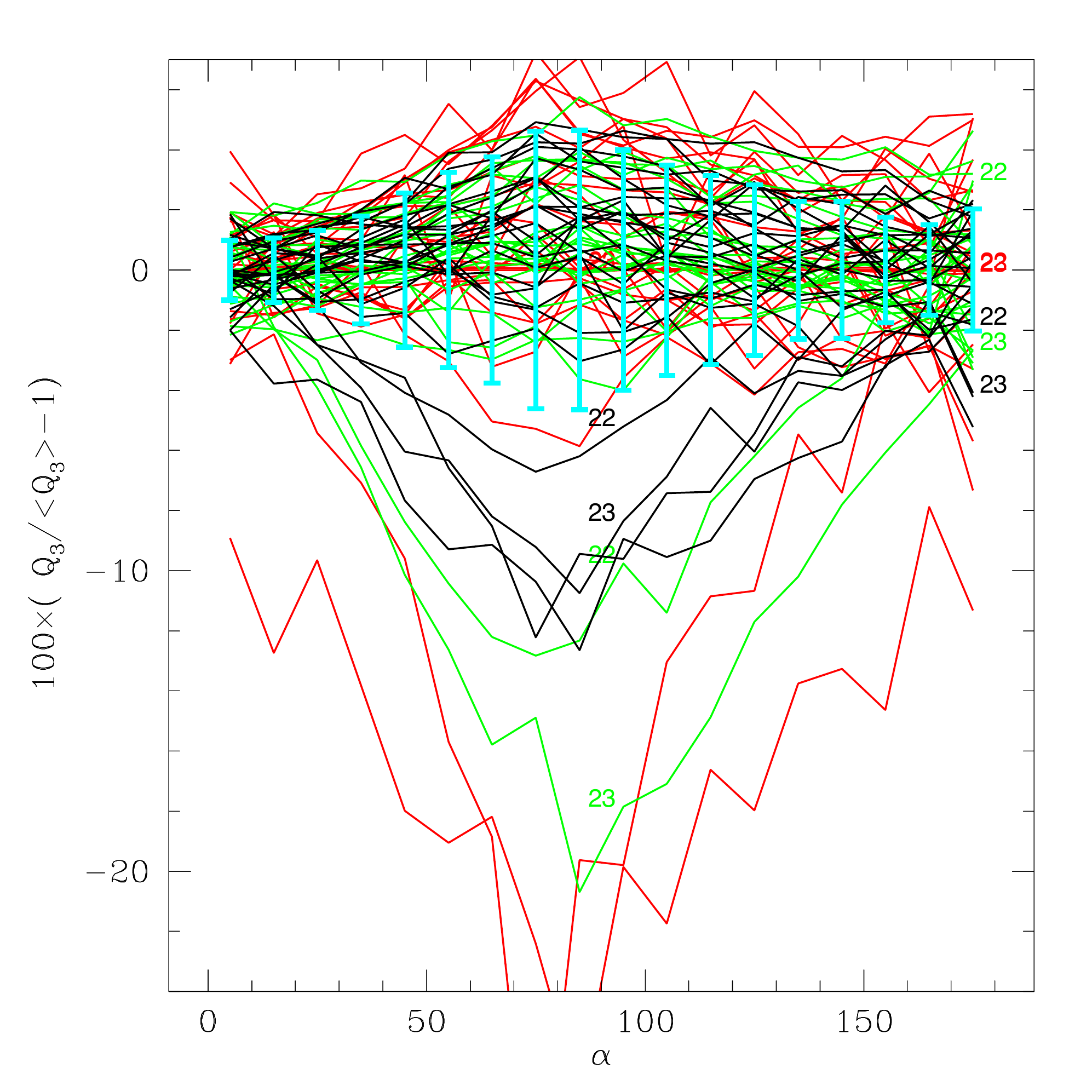}
\caption
{
Clustering statistics measured using Jackknife resamplings of 
bright galaxies from the {\tt bright-all} (black), {\tt bright-far} (red) 
and {\tt bright-ref} (green) samples (see Table~\ref{tab:samples}).
The left-hand panel shows $\xi(s)$ and the right-hand panel 
shows Q$_3(\alpha)$. Outlying correlation functions that were identified 
in the {\tt ref} sample are labelled by the omitted zone number. 
The errorbars correspond to the {\it rms} error from the Jackknife
resamplings of the {\tt bright-all} sample. 
}
\label{fig:ss_impact}
\end{figure*}

The second test we carry out is to compare clustering measurements
made from different volumes. Here we consider bright galaxies
(i.e.\ the half magnitude bin brighter than \Mstar) as this allows us
to cover a wider redshift interval. Again we compare results obtained
from three of the samples listed in Table~\ref{tab:samples}: 
the {\tt bright-all}, the {\tt bright-ref} (the low redshift half of
{\tt bright-all}) and the {\tt bright-far} (the high redshift half of
{\tt bright-all}) samples. The results for the clustering statistics
of the Jackknife resamplings of these galaxy samples are shown in
Fig.~\ref{fig:ss_impact}.  
We have labelled the zones which were identified as outliers in the 
{\tt ref} sample on Fig.~\ref{fig:ss_impact}. Leaving out zone 23
produces an outlier in the measurements made using the 
{\tt bright-ref} sample, as we observed in
Fig.~\ref{fig:galform_impact} already. For the 
{\tt bright-all} and {\tt bright-far} samples, zone 23 is no longer 
an outlier. The removal of other Jackknife zones leads to larger
changes in the measured correlation function. However, the outliers
in these cases are not as dramatic as they were in the case of the {\tt ref}
sample. The right-hand panel of Fig.~\ref{fig:ss_impact} shows the
same statistic for Q$_3(\alpha)$. For the {\tt bright-ref} and 
{\tt bright-all} samples, zone numbers 22 and 23 are outliers, but
again here not as dramatic in  {\tt bright-far} sample. A different
zone is an outlier for the  {\tt bright-far} sample, as happened for
$\xi(s)$.

\begin{figure}
\plotone{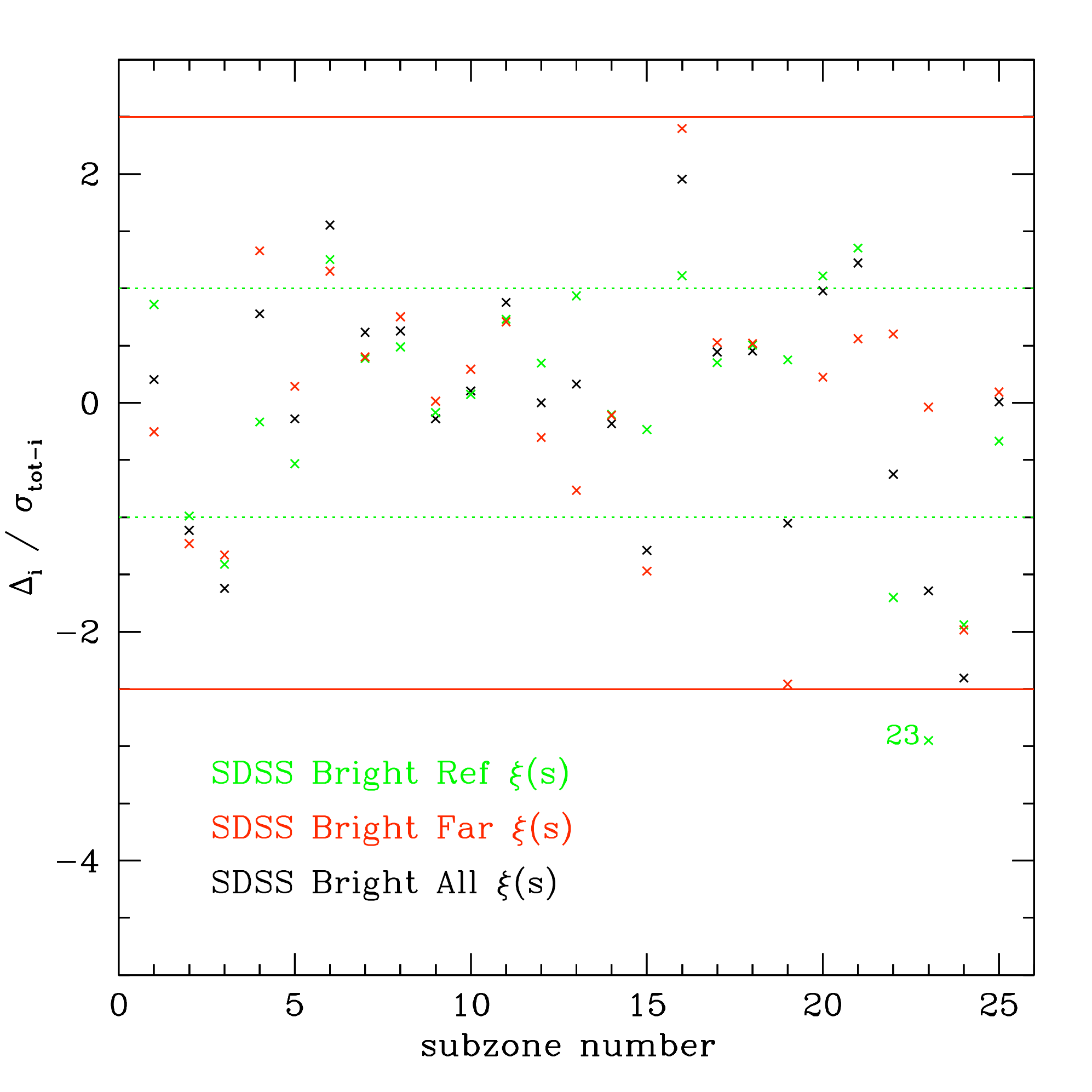}
\caption
{
The JK ensemble statistic for $\xi$ measured using \nsub=25 for 
the {\tt bright-all}, {\tt bright-far}, {\tt bright-ref} volume 
limited samples, plotted in black, red and green respectively.  
As expected, samples with the same type of galaxies but covering 
different volumes do not have the same JK ensemble fluctuation 
but more importantly, the larger the volume the less influential 
any given zone becomes.  
}
\label{fig:JK_fluctuation_ss_impact}
\end{figure}

Again, we quantify the above conclusions by computing the JK ensemble 
fluctuation. We show in Fig.~\ref{fig:JK_fluctuation_ss_impact} the 
JK ensemble fluctuation of $\xi(s)$ for the three samples discussed 
in Fig.~\ref{fig:ss_impact}. As expected, zone 23 is only an outlier 
in terms of this statistic for the {\tt ref} sample, while in the two other
samples, i.e.\ {\tt bright-all} and {\tt bright-far}, the JK ensemble 
fluctuation statistic for $\xi(s)$ points to less extreme results 
for the zones, with {\tt bright-all} being the most ``uniform'' of the three
samples considered: this is due to its volume being twice
that of the two other samples. Finally the most extreme zone
of {\tt bright-all} is a zone which in both {\tt bright-ref} and 
{\tt bright-far} corresponds to $\delta_{JK}^{i} \sim -2.0$ 
in the JK ensemble fluctuation statistic.
Very similar results are found for $Q_3$.

As the zones we use are defined in angle, they cover a wide 
baseline in redshift and many individual structures could contribute
to what we are calling a superstructure. Also, by increasing the volume, 
other structures could be sampled which could have a similar impact 
on the clustering signal. The comparison of the scatter of the Jackknife 
resamplings in the bright sampling volume shows that the omission of 
different zones can lead to outlying clustering measurements. However, 
the departure for a given sub-volume from the rest of population
seems to become less significant the larger the volume.

\subsection{JK ensemble: a practical recipe}

With the JK ensemble fluctuation statistic one can assess whether a
sample has been split into the right number of sub-zones for a
Jackknife error analysis to be valid, as opposed to split into too
many small JK regions. 
The way to proceed is by taking the following steps:
\begin{enumerate}
\item[a.] the size and number of JK regions should be such that there
  is no ``apparent clustering'' in the $\delta_{\rm JK}$ statistic of
  neighbouring zones (i.e.\ neighbouring zones should have
  $\delta_{\rm JK}$ values which are independent from eachother). 
\item[b.] no JK region should present too extreme a value for $\delta_{\rm JK}$
  statistic compared to the others, i.e.\ the associated probability of
  its occurence (as derived from the $\delta_{\rm JK}$ distribution as
  shown in Fig.~5 for $\xi(s)$ with three values of \nsub) should not
  be significantly at odds with the probability of such an event
  actually happening (which for large \nsub\ values can be modelled by a
  Gaussian process with \nsub\ elements). 
\item[c.] The probability threshold recommended by the present work is
  $\sim 3$\% (i.e.\ about 2-$\sigma$ if Gaussian distributed). This
  corresponds to $\delta_{\rm JK}=-2.5$ for $\xi(s)$ with \nsub=27,
  and $\delta_{\rm JK} \simeq -2$ for larger \nsub\ values (according
  to Fig.~5). 
\end{enumerate}
It is not always possible to find an appropriate number of JK
subsamples into which the survey should be split which satisfies both
conditions (a) and (b), unless one is reduced to use far too few
subsamples from which a statistical inference can be made. In the
limit \nsub=1, both conditions are automatically satisfied, but no
errors can be inferred. For example in our analysis of the
SDSS \Mstar\ sample ({\tt ref} sample in Table~\ref{tab:samples}),
\nsub=25 is not appropriate, but at least better than \nsub=225. 
The JK ensemble fluctuation statistic provides a quantitative
measure of the limitation of the error analysis and an indication of
the need to proceed with further checks before statistically
interpreting the results. For example, do the outliers agree with the
expectations from simulations? In summary, when conditions (a) and (b)
cannot be satisfied, we should use statistical inference with greater
care than simply assuming that a comprehensive (but inappropriate)
error analysis has solved the problem entierly.

\begin{figure}
\plotone{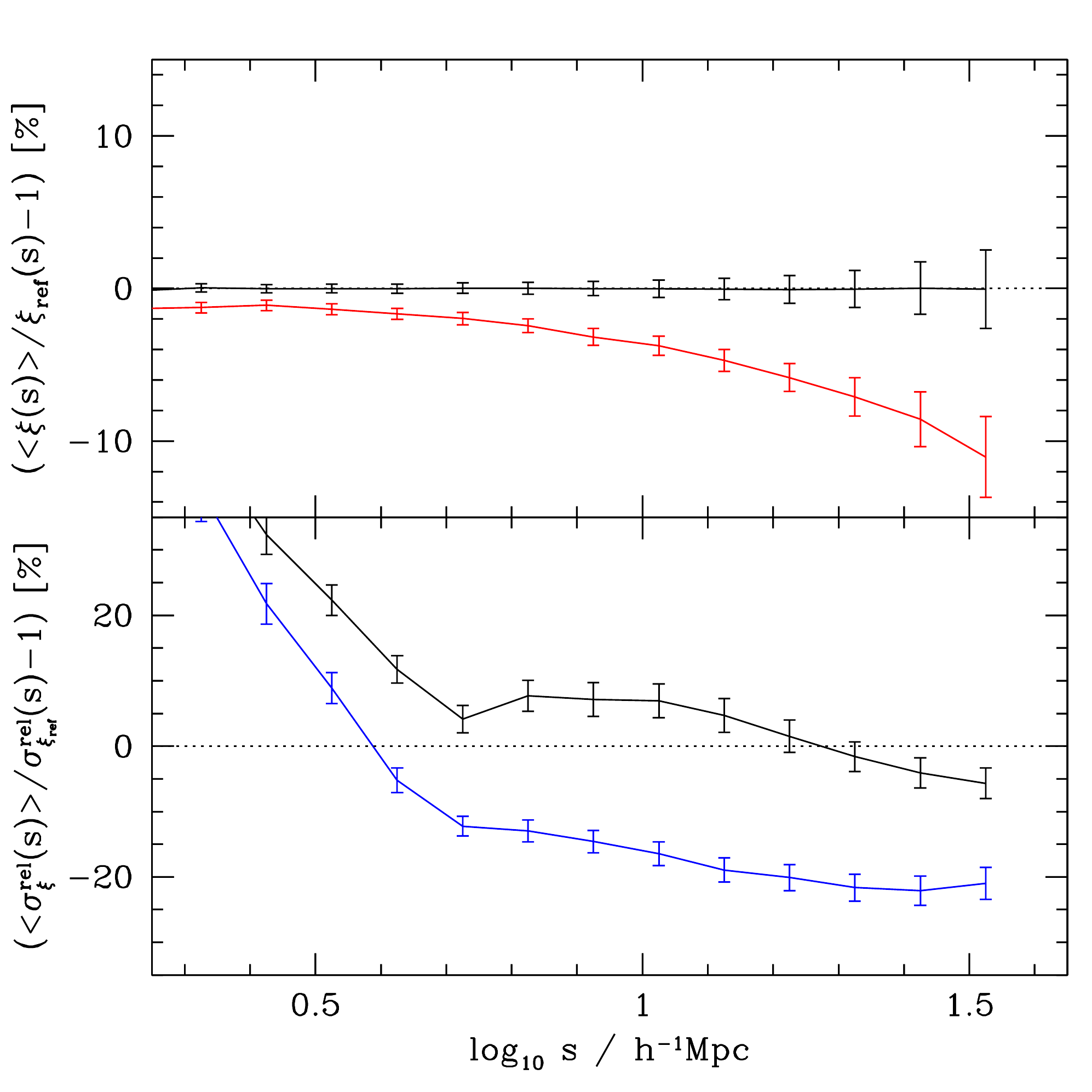}
\caption
{
{\bf Top}: The percentage deviation in the 2-pt correlation function over
the N-body ensemble of realisations.
The mean result for all the JK resamplings is shown in black, while the red
curve shows the result of considering only the outlying
regions, which are defined as those for which $|\delta_{\rm JK}|> 2.5$.
Errorbars show the error on the mean.
{\bf Bottom}: The percentage difference in the relative error on the
2-pt correlation function compared to the N-body ensemble of
realisations. As in the top panel, the black line shows the mean result
for the ensemble of JK resamplings, while the blue
curve presents the result of estimating the 2-pt clustering error
after discarding the outlying JK region from the error analysis, with
errorbars showing the error on the mean.
The top panel shows that discarding the outlying regions affects
systematically the overall clustering amplitude on the scales of
interest (from $\sim2$ to $\sim10$ per cent between 3 and
30$\Mpc$). The bottom panel shows that ignoring the outliers just for
the error analysis introduces a systematic underestimate of $\sim10$
to $\sim20$ per cent on those same scales, while the ensemble of JK
resamplings reproduces fairly accurately the results from the ensemble
of N-body simulations (to within a few percent).
Further details are given in~\S\ref{sec:ignore_outliers}.
}
\label{fig:ignore_outlier}
\end{figure}

\subsection{Should one ignore the outliers in the analysis?}
\label{sec:ignore_outliers}

An interesting question to ask is: "Is it better to ignore an outlying
region altogether when estimating a clustering statistic and its
associated error?" 
From a Bayesian point of view, such a ``massaging'' of the data set is
clearly unattractive, but because it has regularly happened in the past
(e.g. Zehavi \etal\ 2002, 2005, 2011 for 2-pt clustering statistics of
SDSS \Mstar\ samples) we investigate here what consequences this can
have. 

A proper analysis to answer the above question would require the
recalculation of all clustering statistics and discarding from the
start all JK regions identified as outliers. Indeed, the way our JK
ensemble fluctuation works is such that if JK region $i$ is an
outlier, then all other JK estimates contain region $i$. Hence
discarding JK region $i$ requires the re-estimation of the clustering
signal using only the remaining \nsub-1 JK regions. Such a calculation
is unfortunately beyond the scope of this paper\footnote{The CPU time
  required to do the full analysis is close to equivalent to what was
  required in Norberg \etal\ (2009).}, so instead we propose the
following two tests:
\begin{itemize}
\item[a.] the mean clustering signal from outlying regions should be
  comparable to the mean clustering signal from the ensemble of
  simulations. If not, then ignoring the outlying regions will
  influence the overall amplitude of the clustering signal, which
  clearly is unwanted. 
\item[b.] if one cannot ignore the outlying JK regions in the estimate of
  the clustering statistic, is it valid to ignore them in the error
  estimate of the clustering signal? The point here is to understand
  whether excessive fluctuation due to a ``single'' region is better
  discarded in the error analysis. 
\end{itemize}
These two tests are minimal conditions to be satisfied for discarding
outlying regions from the analysis, where outlying region is defined
here by $|\delta_{\rm JK}|>2.5$, appropriate for a $\sim 2$-$\sigma$
cut for \nsub=27. 

To answer test (a), the top panel of Fig.~\ref{fig:ignore_outlier}
displays the relative effect on the 2-point clustering signal of
discarding the outlying regions (red) or using the ensemble of JK
resamlpings (black). In both cases the results are compared to the
clustering signal from the ensemble of N-body simulations
($\xi_{ref}(s)$, with error $\sigma_{\xi_{ref}}(s)$).
The errorbars show the error on the mean, which is the relevant
quantity for the present analysis as we are interested in the mean
trend and any deviations from its expected value (as shown by the
dotted horizontal line). To avoid new estimates to be calculated, the
red curve shows the results from averaging the clustering
signal in all the outlying regions, selected from each simulation via
the condition $|\delta_{\rm JK}|>2.5$. We point out that we had to
ignore volumes with two or more outliers ($\sim10$ per
cent of the volumes), as in those cases the clustering signal of an
outlying JK region is not necessarily representative of the clustering
of the volume with all outlying regions removed.
The top panel of Fig.~\ref{fig:ignore_outlier} clearly shows that 
the outlying JK regions have a systematically different clustering
signal from the simulation ensemble, hence ignoring them in any
clustering analysis would result in systematically low clustering
amplitudes. We additionally note that there is scale dependence in
this bias, changing from $\sim2$ to $\sim10$ per cent between 3 and 30
$\Mpc$. 

To answer test (b), the bottom panel of Fig.~\ref{fig:ignore_outlier}
displays the percentage difference in the relative error on the
2-point clustering signal of discarding the outlying regions (blue) or
using the ensemble of JK resamplings (black). In both cases the
results are compared to the relative clustering error from the ensemble of
N-body simulations, i.e. $\sigma_{\xi_{ref}}^{rel}(s)$. As for the top
panel, the errorbars indicate the error on the mean. 
Since test (a) showed that it is
incorrect to ignore the outlying JK region for the estimate of
$\xi(s)$ and to avoid calculating new clustering estimates, the
blue curve shows in fact the results from averaging the relative error
on $\xi(s)$ ignoring the contribution from all the outlying regions,
selected from each simulation via the condition 
$|\delta_{\rm JK}|>2.5$. Yet again, only simulation volumes with one
single outlying region were considered, as only for these volumes can
we relate the excess clustering to one JK region (which does not
require any reestimation of the clustering statistics).  
The bottom panel of Fig.~\ref{fig:ignore_outlier} 
shows that, while the relative error from the ensemble of JK
resamplings is within a few per cent of the relative error from the
simulation ensemble, the relative error estimated from ignoring the
outlying regions is underestimated by typically 15 to 20 per cent on
scales above $\sim5\Mpc$. It is worth noting that the error on scales
smaller than a few $\Mpc$ is, as 
already noted in Norberg \etal\ (2009), overestimated by a significant
amount when using JK resamplings compared to the simulation ensemble.

Hence using our series of N-body simulations we have shown in this
section that not including the outlying JK region in the estimate of
the clustering statistic and its errors significantly affects the
results and hence should be avoided.

\begin{figure*}
\plotfull{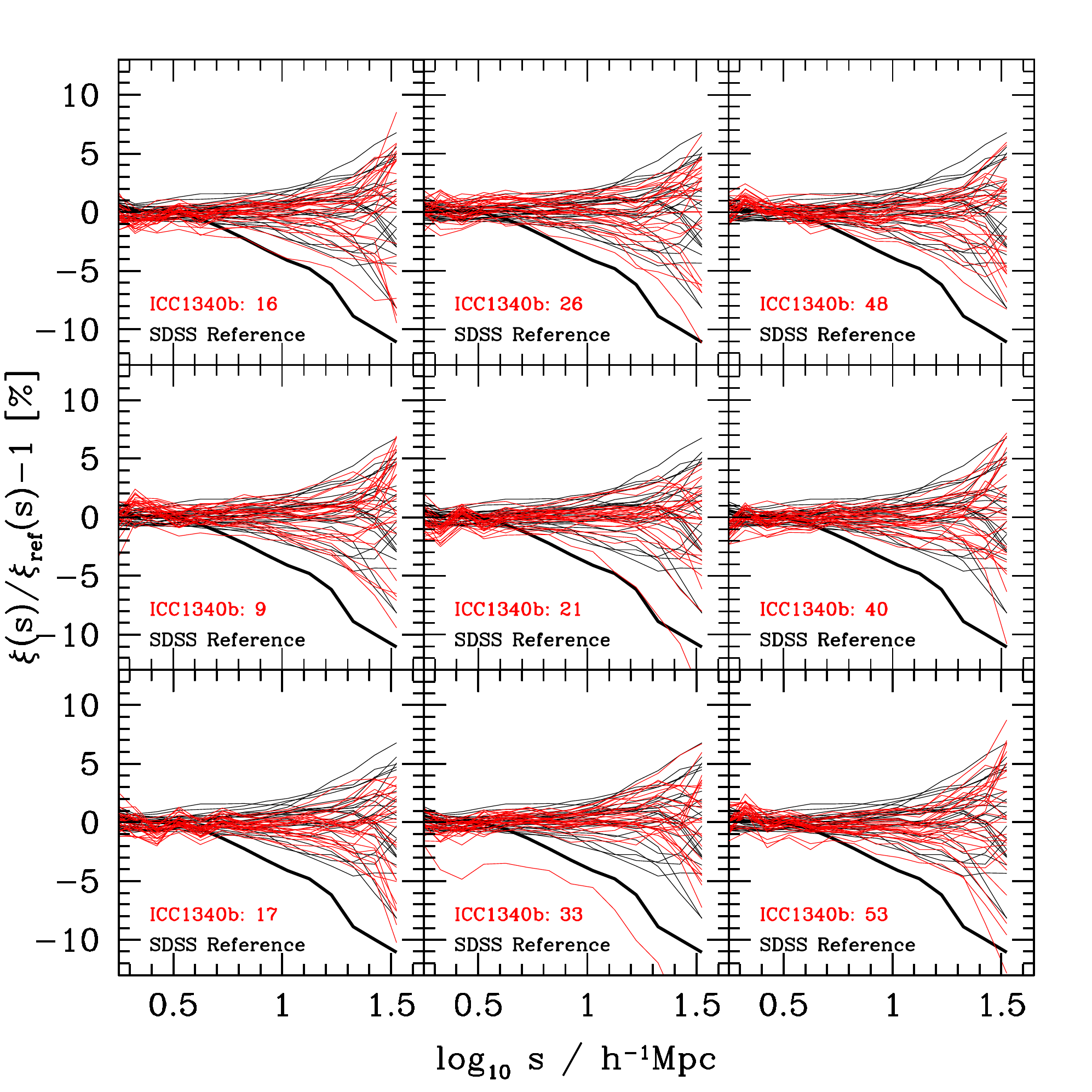}
\caption
{
The JK resampling fluctuation for the two-point correlation 
function in nine (one per panel) randomly chosen {\tt L-BASICC}
\LCDM\ simulations that are all statistically similar and from 
which a volume comparable to a \Mstar\ SDSS sample has been extracted 
(red lines). The equivalent statistic for the SDSS data is shown 
using black lines and is reproduced in each panel. The thick 
black line shows the clustering measured on omitting zone 23. 
}
\label{fig:lcdm_sim}
\end{figure*}

\section{Is the SDSS \Mstar\ sample compatible with $\Lambda$CDM?}
\label{sec:lcdm_sim}

Having identified an unusual overdensity of \Mstar\ galaxies in the 
SDSS, it is natural to ask if such a structure is expected in the 
\LCDM\ cosmology. We address this question using a suite of large
volume N-body simulations by Angulo \etal\ (2008). The same
calculations were used in the evaluation of internal error estimation
schemes by Norberg \etal\ (2009).  

The {\tt L-BASICC} simulation ensemble of Angulo \etal\ (2008) comprises 
50 moderate resolution runs, each representing the matter distribution 
using $448^3$ particles of mass $1.85\times10^{12}\,h^{-1}\,M_\odot$ in a
box of side $1340h^{-1}$Mpc. Each L-BASICC simulation was evolved from
a different realization of a Gaussian density field set up at
$z=63$, using the following cosmological parameters
%The adopted WMAP1 cosmological parameters are broadly consistent
%with recent data from the cosmic microwave background and the power
%spectrum of galaxy clustering  
%(e.g.\ Sanchez \etal\ 2006, 2009): 
$\Omega_{\rm M}=0.25$, $\Omega_{\Lambda}=0.75$, $\sigma_{8}=0.9$,
$n=1$, $w=-1$ and $h=H_{0}/(100\,{\rm km\,s}^{-1}{\rm Mpc}^{-1})=0.73$.   
The combination of a large number of independent realizations and the
huge simulation volume make the L-BASICC ensemble ideal for searching 
for unusual structures (see also Yaryura, Baugh \& Angulo 2010).

Following the method set out in Norberg \etal\ (2009), we extract from 
each simulation cube two cubical sub-volumes of $380~\Mpc$ on a side,  
which are separated by at least $\sim 500~\Mpc$. These pairs of volumes 
are correlated because they come from the same simulation box. However, 
in practice, because of the wide separation of the subvolumes and the 
low amplitude of the power at these wavelengths, we can treat them as 
being independent. Hence, with this procedure we construct 100 mock
catalogues each of which is fully independent of 98 of the others
(as these come from different simulation boxes) and is essentially 
independent of the other subvolume taken from the same simulation.  

The redshift space distortion of clustering is not modelled using the
distant observer approximation within the extracted region. 
Instead we place the observer at the centre of the simulation 
volume and extract a region that is at a distance comparable to the SDSS
\Mstar\ sample, i.e.\ $\sim 300\Mpc$ away. 
The cubical region that we extract is somewhat bigger than the SDSS 
\Mstar\ sample. This is to provide a spatial buffer, as peculiar 
velocities can displace particles in either direction along the 
line of sight, and particles can be moved into as well as out of 
the volume of interest. Finally, we randomly dilute the number of
dark matter particles in each data set to match the number density of
the \Mstar\ galaxy sample of $3.7 \times 10^{-3}~\Mpccube$, which 
mimics the discreteness or shot noise level in the SDSS \Mstar\ volume 
limited sample.
Following Norberg \etal\ (2009), we do not attempt to model a particular
galaxy sample and survey geometry in detail, but simply to cover a 
comparable volume with the same number density of objects. Our aim here 
is to make a generic comparison: are enough large structures
expected in the \LCDM\ cosmology to account for the outliers we 
have seen in 1/25$^{\rm th}$ of a volume limited \Mstar\ SDSS 
galaxy sample?

We repeat the Jackknife resampling analysis we carried out earlier for the
SDSS data samples, but for the {\tt L-BASICC} dark matter catalogues
and present the 2-point clustering estimates in Fig.~\ref{fig:lcdm_sim}. 
Each panel shows results for a different subvolume from the ensemble 
(in red with the subsample number indicated in the panel legend) and 
the corresponding result for the SDSS {\tt ref} sample (in black and 
replicated in all panels). Each line corresponds to the fluctuation 
of $\xi(s)$ measured from one of the N$_{\rm sub}$ Jackknife resamplings 
around the measurement from the corresponding full volume limited sample.

The L-BASICC regions we analyze have a volume equal to that of the 
{\tt bright-all} sample, which is twice the size of the {\tt ref} sample, 
so we expect the scatter in the Jackknife resamplings for the simulations 
to be smaller than for the {\tt ref} sample. We found a similar 
result in Fig.~\ref{fig:ss_impact} with the scatter from the  
{\tt bright-all} sample being systematically smaller than either 
of its two components, i.e.\ {\tt bright-far} and {\tt bright-far}.

Mainly for this reason and for slight mismatches between the 
simulation and the SDSS data (see Norberg \etal\ 2009 for a
detailed discussion), the scatter measured in the simulation results
should be considered as a lower limit compared to that displayed by
the SDSS data in the {\tt ref} sample,
or better matched to the {\tt bright-all} sample (based on simple
volume arguments). 
The main conclusion we can draw
from Fig.~\ref{fig:lcdm_sim} is that outliers comparable to or even 
more extreme than that produced by zone 23 in the data 
are reasonably common in the \LCDM\ cosmology. In approximately 
one third of the randomly chosen cases, the scatter in the simulation 
Jackknife resamplings is comparable to what is seen in our reference 
volume limited sample. Hence the structure in zone 23 does not 
present a problem for \LCDM\ .

\section{Superstructures in 2\lowercase{d}FGRS \& SDSS}
\label{sec:SDSS_2dFGRS}

\begin{figure}
\plotone{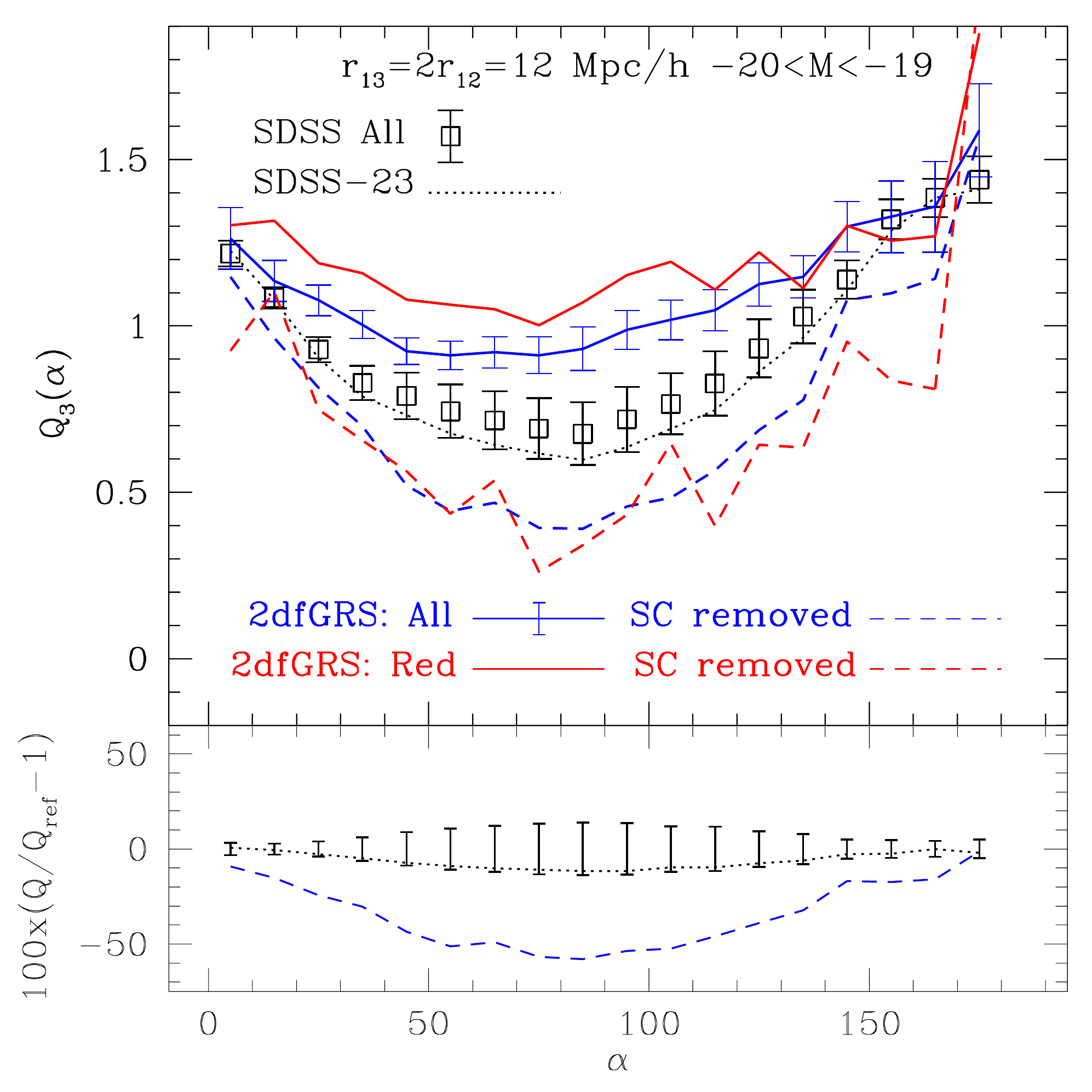}
\caption
{Comparison of Q$_3(\alpha)$ estimates from the 2dFGRS (blue continuous
line with errorbars) and SDSS (open squares with errorbars).
SDSS errorbars are from the $N_{\rm sub}=25$ JK subsamples. 2dFGRS errors
are from the dispersion of 22 galaxy mocks that match 2dFGRS
2-point correlations (see Gazta\~naga \etal\ 2005 for details).
The dashed blue line corresponds to the 2dFGRS measurement after removing 
the 2 largest superclusters. The dotted line corresponds to result from the 
SDSS after leaving out JK subsample 23 (ie all minus outlier subregion 23 in 
Fig.\ref{fig:sdss_quilt_zoom}). Red lines show the corresponding results
from the 2dFGRS red selected sample, which ought to provide a better comparison 
with the r-band SDSS selected galaxies. Errorbars in the bottom panel show the 
relative JK rms fluctuations in SDSS (same as in Fig.~\ref{fig:ratio_stats} 
but scaled by usual JK factor $\sqrt{N_{\rm sub}}= 5$). The dotted line 
corresponds to results after leaving out SDSS JK subsample 23. The blue dashed 
line shows the relative fluctuation for the 2dFGRS sample after excluding 
the superstructures. }
\label{fig:clustering_comparison}
\end{figure}

Previous studies which estimated the higher-order clustering
statistics from the 2dFGRS (e.g.\ Baugh \etal\ 2004; Croton
\etal\ 2004, 2007; Gazta\~naga \etal\ 2005) and SDSS surveys 
(e.g.\ Nichol \etal\ 2006;
Mcbride \etal\ 2011)
found that the presence of one or two superstructures   
had a strong influence on the interpretation of the measurements 
for their \Mstar\ samples, which are all similar in their definitions
to the one considered here, covering roughly the same redshift range,
i.e.\ $0.05 \lesssim z \lesssim 0.11$ and sampling mostly
\Mstar\ galaxies.
In some work, clustering statistics were presented both for full
samples and for samples  
from which the volume containing the superstructure had been cut out, referred 
to as the ``with'' and ``without'' supercluster results. This practice 
was merely intended to illustrate the impact of these structures on the 
measurements, rather than to advocate one or the other as being the definitive 
or correct answer. An example of this practice is shown for the 2dFGRS
in Fig.~\ref{fig:clustering_comparison}, which  
shows the ``with'' and ``without'' 
measurements 
of the 3-point 
function for the \Mstar\ volume limited sample, as first presented in 
Gazta\~naga \etal\ (2005). At that time, because of the substantial
difference between these two measurements,  
the \Mstar\ sample results were considered unreliable. Therefore the
analysis and interpretation focused mostly on larger volume limited samples,
corresponding to brighter galaxies. Due to a combination of increased volume 
and the way in which brighter galaxies sample or weight structures (compared 
to \Mstar\ galaxies) these volume limited samples seemed to be less affected 
by any particular superstructure.

There is one subtle, but important difference between these previous analyses
and the one presented in this paper. Here an objective, blind approach
is taken,  
while in the earlier works, a bespoke supercluster mask was
constructed, starting from the location of high peaks in the 3-D
density field and then masking all galaxies within a radius of
$25\Mpc$ from the centre of the peak (see Baugh \etal\ 2004 and Croton
\etal\ 2004 for details). This method actually results in a more
significant difference between the measurements ``with'' and
``without'' the superclusters. It is therefore natural to revisit
these results and compare them with the more objective method
considered here.

In Fig.~\ref{fig:clustering_comparison} we show a comparison between
Q$_3(\alpha)$ estimated from the 2dFGRS and SDSS \Mstar\ samples.  
The 2dFGRS clustering estimates are from Gazta\~naga \etal\ (2005),
with two sets of measurements plotted: the with superclusters measurement 
(continuous lines) and the without superclusters measurement (dashed lines).
The SDSS DR7 measurements from the {\tt ref} sample (squares with errorbars) are
new and fall consistently between the results from the 2dFGRS analysis. 
It is reassuring to see how clustering measurements for 3-point
statistics are reproducible: they come from different parts
of the sky, are obtained using different telescopes/cameras and, most
importantly, are based on slightly different galaxy selections. 
To attempt to remove the difference in selection between the 
2dFGRS and SDSS samples, we also show the results for the 2dFGRS, both 
with and without superclusters, for a red selected subsample 
(see Gazta\~naga \etal\ 2005 for the exact definition of their red
selected 2dFGRS \Mstar\ sample).
This should be more comparable to the selection in the SDSS, which is 
done in the $r-$band. The results are noisier because of the smaller
number of galaxies retained, but are similar to the measurements from 
the full, blue selected 2dFGRS sample. 

In the bottom panel of Fig.~\ref{fig:clustering_comparison}, the dashed (blue)
line indicates the relative difference in Q$_3(\alpha)$ (in percent)
measured from the 2dFGRS when we compare results with and without the
superclusters. The effect seen in the 2dFGRS is much larger than the
analogous result in Fig.~\ref{fig:ratio_stats} for the SDSS. The SDSS
measurement is characterized by the errorbars in the bottom panel of
Fig.~\ref{fig:clustering_comparison}, scaled by the usual JK factor
$\sqrt{N_{\rm sub}}= 5$ to show the variance in the full sample (as
opposed to the variance in the JK-subsample displayed in
Fig.~\ref{fig:ratio_stats}). The shift in the measurement of 
$Q_3$ from the SDSS on leaving out the outlier subregion 23 is shown
by the dotted line. For the 2dFGRS, removing the supercluster from the
clustering analysis produces a change of up to $\sim 50$ per cent in
Q$_3(\alpha)$, while the corresponding number for the SDSS using the
more objective method of JK resampling introduced here is 
$\sim 12$ per cent. These differences reflect both that the 2dFGRS 
is $\sim 5$ times smaller than SDSS and also the different ways in
which outlying superstructures are defined in each case. Also note how
the SDSS measurement on leaving out the outlier subregion is still
within the JK errors while the 2dFGRS measurement without
superclusters is outside the errors (which came from an ensemble of 22
mock catalogues). 

What have we learnt from this? From our new outlier analysis using
Jackknife resampling and for the particular statistics (and scales)
we consider, the impact of large scale inhomogeneities in the SDSS
\Mstar\ sample can be as large as $\sim 12\%$ in Q$_3(\alpha)$ on
scales of $12-14\Mpc$. 
We can and should use this information to test models,
both in terms of interpreting a fit to the data and in terms of the  
outlier probability in realizations of a given model. These outlier
probabilities 
are now significantly better defined than before, as is clear from 
the bottom panel of Fig.~\ref{fig:clustering_comparison} through the difference 
between the 2dFGRS estimate ``with'' and ``without'' superclusters and the 
statistical Jackknife resampling method. More importantly, as shown in the 
previous section, we can apply the same objective algorithm
(i.e.\ the Jackknife resampling method) to both data and simulations
enabling statistically robust conclusions to be drawn. The previous
method applied to the 2dFGRS, based on removing superstructures, is
subjective and significantly harder to transfer from data to mocks. 
It is now clear from the comparison between the 2dFGRS and SDSS results 
in Fig.~\ref{fig:clustering_comparison} that the error bars derived 
from mock 2dFGRS catalogues did not capture the full variance in that
sample. 
This can be better understood if one accounts for the various
limitations of the mocks, i.e.\ (a) we have only 22 non-overlapping (but
still not independent) mocks drawn from the same Hubble Volume
simulation (Evrard \etal\ 2002); (b) the mocks are only constrained to
reproduce the two-point 2dFGRS galaxy clustering (Cole \etal\ 1998);
(c) errors on clustering statistics depend directly on their higher
order moments 
(e.g.\ Bernardeau \etal\ 2002).    
This was already 
noted by Gaztanaga \etal\ (2005) who considered this sample to be unreliable 
because of the effect of the superclusters. 
The situation is different for the SDSS 
analysis presented here (in Fig.~\ref{fig:clustering_comparison}) where, despite 
the presence of the outlier region 23 (i.e.\ see Fig.\ref{fig:sdss_quilt_zoom}), 
we can be more confident of the error analysis because we have a systematic
way to quantify the impact of outliers.

\section{Summary}
\label{sec:conclusion}

This paper presents a new method to quantify the robustness of error
estimates for clustering statistics. The approach is an extension of
the Jackknife technique and uses two new clustering diagnostics 
to quantify the distribution of clustering measurements from 
different resamplings: the JK resampling fluctuation (first 
shown in Fig.~\ref{fig:ratio_stats}), and the JK ensemble 
fluctuation (first shown in Fig.~\ref{fig:JK_fluctuation}). 
The main features of the method can be summarized as follows: 
\begin{itemize}

\item The technique provides an objective way of finding large coherent
  structures. As the Jackknife zones are set up a priori, this is a 
  blind test for the presences of superstructures, based on their 
  impact on the measured clustering statistics. This is an 
  improvement over earlier work in which ``unusual'' regions were omitted 
  after making the clustering measurements, without any firm guidance 
  as to what volume should be excluded. 

\item Our approach provides a quantitative way of determining the
  appropriate size of the Jackknife zones to be used in the error
  analysis of the clustering signal. 
  It follows these three key ingredients:
  \begin{enumerate}
\item[a.] the size and number of JK regions should be such that there
  is no ``apparent clustering'' in the $\delta_{\rm JK}$ statistic of
  neighbouring zones (i.e.\ neighbouring zones should have
  $\delta_{\rm JK}$ values which are independent from eachother).
\item[b.] no JK region should present too extreme a $\delta_{\rm JK}$
  statistic compared to the others, i.e.\ the associated probability of
  its occurence (as derived from the $\delta_{\rm JK}$ distribution as
  shown in Fig.~\ref{fig:JK_new} for $\xi(s)$ with three values of
  \nsub) should not be significantly at odds with the probability of
  such an event actually happening (which for large \nsub\ values can
  be modelled by a Gaussian process with \nsub\ elements). 
\item[c.] The probability threshold recommended by the present work is
  $\sim 3$\% (i.e.\ about 2-$\sigma$ if Gaussian distributed). This
  corresponds to $\delta_{\rm JK}=-2.5$ for $\xi(s)$ with \nsub=27,
  and $\delta_{\rm JK} \simeq -2$ for larger \nsub\ values (according
  to Fig.~\ref{fig:JK_new}). 
  \end{enumerate}

\item The new statistics we have introduced can be used to compare
  observations to models. As shown in \S\ref{sec:lcdm_sim}, we can
  check in a 
  straight forward and objective  way whether or not simulations
  produce similar outlying clustering measurements as those seen in
  the data. Again, this is a significant improvement over the ad-hoc
  approach of computing the correlation functions ``with'' and
  ``without'' the superstructrures. 

\item The statistics offer a quantitative way to study the influence
  of large coherent structures on the measured correlation function,
  even if the structures dominate the clustering signal. As shown in
  \S\ref{sec:SDSS_2dFGRS}, with these new tools we can quantify how
  inhomogeneities affect our measurements and the reliability of our
  error estimates.

\end{itemize}

In the standard application of the Jackknife method to galaxy surveys, 
the dataset is divided spatially into zones. Our procedure overcomes 
one of the long standing problems of Jackknife error estimation: 
how many zones should the data be split up into? A large number of 
zones implies a large number of resamplings of the dataset. To 
improve the stability of the inversion of the covariance martix, 
it is important to have as large as possible a number of resamplings 
(e.g.\ Hartlap \etal\ 2007). 
This is a vague concept, as if the Jackknife zones are made too small, 
the resamplings will be essentially the same, with a very small 
change in the volume covered. Our approach uses the Jackknife 
ensemble fluctuation to determine the appropriate number of zones. 
This statistic allows us to identify the omission of a zone as leading 
to an outlying clustering measurement. We have argued that such 
outliers should be due to the structure contained within one Jackknife 
zone, rather than several contiguous zones. In our illustration using the 
\Mstar\ SDSS volume limited sample, this suggested that \nsub=25 is
a more robust number of zones to use in the Jackknife method than
\nsub=225. 

The approach we have set out in this paper, along with the comparison
between different error estimation methods presented in Norberg
\etal\ (2009), provides a robust and objective presciption for the
estimation of galaxy correlation functions and their associated
errors, which can be applied to any type of forthcoming survey, both
spectroscopic and photometric.  

\section*{Acknowledgements}

We acknowledge the referee for delivering a very useful report.
PN wishes to acknowledge numerous stimulating discussions taking place
in the East Tower at the IfA and the kind use of the distributed
computing resources at the IfA.  
PN has been supported by the IfA's STFC rolling grant, by a
Royal Society University Research Fellowship and by an ERC StG Grant
(DEGAS-259586). 
EG acknowledges support from Spanish Ministerio de Ciencia y Tecnologia
(MEC), projects AYA2009-13936 and Consolider-Ingenio CSD2007-00060,
and research project 2009SGR1398  from  Generalitat de Catalunya.
%
%CMB is supported by a Royal Society University Research Fellowship. 
%
%DC acknowledges the financial support from NSF grant AST00-71048.
DC acknowledges receipt of a QEII Fellowship awarded by the Australian
government. 

Funding for the Sloan Digital Sky Survey (SDSS) and SDSS-II has been
provided by the Alfred P. Sloan Foundation, the Participating
Institutions, the National Science Foundation, the U.S. Department of
Energy, the National Aeronautics and Space Administration, the
Japanese Monbukagakusho, and the Max Planck Society, and the Higher
Education Funding Council for England. The SDSS Web site is
http://www.sdss.org/. 

The SDSS is managed by the Astrophysical Research Consortium (ARC) for
the Participating Institutions. The Participating Institutions are the
American Museum of Natural History, Astrophysical Institute Potsdam,
University of Basel, University of Cambridge, Case Western Reserve
University, The University of Chicago, Drexel University, Fermilab,
the Institute for Advanced Study, the Japan Participation Group, The
Johns Hopkins University, the Joint Institute for Nuclear
Astrophysics, the Kavli Institute for Particle Astrophysics and
Cosmology, the Korean Scientist Group, the Chinese Academy of Sciences
(LAMOST), Los Alamos National Laboratory, the Max-Planck-Institute for
Astronomy (MPIA), the Max-Planck-Institute for Astrophysics (MPA), New
Mexico State University, Ohio State University, University of
Pittsburgh, University of Portsmouth, Princeton University, the United
States Naval Observatory, and the University of Washington. 

\appendix

\section{SQL queries}
\label{sec:appendix_query}

In this appendix we list the SQL queries and data files used
to generate the galaxy catalogues and to construct the imaging and 
spectroscopic completeness masks used in the paper. 

\subsection{Galaxy Catalogue} 

The CasJobs SQL query used to generate the input galaxy catalogue from
which the spectroscopic galaxy catalogue is derived is:\\

{\tt
\noindent
SELECT ... \\
INTO ... \\ % v4_drX_photoprimary_targetID_v0   -- v4_drX lists the galaxies
FROM PhotoPrimary po \\
LEFT OUTER JOIN Target t on t.bestObjID = po.objid \\
LEFT OUTER JOIN SpecObj so on po.specObjID = so.specObjID \\
where po.primtarget\&(64|128|256)!=0 and po.status\&(16384)!=0 \\
      and t.targetID > 0 \\
}

\noindent
This query accounts for galaxies with and without redshifts, but which
were intended for spectroscopic targeting. Only galaxies
whose redshift satisfies the GAMA (Driver \etal\ 2009, 2010) selection
criteria for a good redshift (see Baldry \etal\ 2010
for the exact definition) are used in the clustering analysis.

\subsection{Imaging and Spectroscopic Mask} 

To generate the imaging and spectroscopic masks, the following CasJobs SQL
query can be used :\\ 

\noindent
{\tt
SELECT \\
run,field,raMin,raMax,decMin,decMax,\\
rerun,camcol,skyVersion \\
FROM field \\
}

\noindent
together with these two DR7 files:

\noindent
{\tt
http://www.sdss.org/dr7/coverage/tsChunk.dr7.best.par
}

\noindent
{\tt
http://www.sdss.org/dr7/coverage/maindr72spectro.par
}

\noindent
listing the imaging coverage and the position of the spectroscopic
tiles respectively. It is also necessary to include the information
about additional ``holes'' in SDSS DR7 coverage, as listed on

\noindent
{\tt
http://www.sdss.org/dr7/coverage/holes.html .
}

\noindent
The masks, quilts and galaxy catalogues can be made available upon
request by contacting lead author.


\begin{thebibliography}{99} 

\bibitem [\protect\citename{XX} XX]{Abazajian09} 
Abazajian K.N., \etal, 2009, ApJS, 182, 543
\bibitem [\protect\citename{XX} XX]{Angulo08} 
Angulo R.E., Baugh C.M., Frenk C.S., Lacey C.G., 2008, MNRAS, 383, 755

\bibitem [\protect\citename{Baldry \etal} 2010]{Baldry10}
Baldry I.K., \etal, 2010, MNRAS, 404, 86
\bibitem [\protect\citename{Baugh \etal} 1995]{Baugh95} 
Baugh C.M., Gazta\~naga E. \& Efstathiou G., 1995, MNRAS, 274, 1049
\bibitem [\protect\citename{Baugh \etal} 2004]{Baugh04} 
Baugh C.M., \etal, 2004, MNRAS, 351, L44
%Estimating Fixed-Frame Galaxy Magnitudes in the Sloan Digital Sky Survey
%{\bf
\bibitem [\protect\citename{XX} XX]{Bernardeau02} 
Bernardeau F., Colombi S., Gazta\~naga E., \& Scoccimarro R., 2002,
Phys. Rep. 367, 1
%}
\bibitem [\protect\citename{XX} XX]{Blanton03a} 
Blanton M.R., \etal, 2003, AJ, 125, 2348 
%The Galaxy Luminosity Function and Luminosity Density at Redshift z = 0.1
\bibitem [\protect\citename{XX} XX]{Blanton03b} 
Blanton M.R., \etal, 2003, ApJ, 592, 819
%%The Broadband Optical Properties of Galaxies with Redshifts 0.02<z<0.22
%\bibitem [\protect\citename{XX} XX]{Blanton03c} 
%Blanton M.R., \etal, 2003, ApJ, 594, 186

%{\bf
\bibitem [\protect\citename{Cole \etal} 1998]{Cole98} 
Cole S., Hatton S., Weinberg D.H., Frenk C.S., 1998, MNRAS 300, 945
%}
\bibitem [\protect\citename{Cole \etal} 2005]{Cole05} 
Cole S., \etal, 2005, MNRAS 362, 505
\bibitem [\protect\citename{Colless \etal} 2001]{Colless01} 
Colless M., \etal, 2001, MNRAS, 328, 1039
\bibitem [\protect\citename{Colless \etal} 2003]{Colless03} 
Colless M., \etal, 2003, astro-ph/0306581
%{\bf
\bibitem [\protect\citename{Croft \etal} 2004]{Croft99} 
Croft R.A.C., Dalton G.B. \& Efstathiou G., 1999, MNRAS 305, 547	
%}
\bibitem [\protect\citename{Croton \etal} 2004]{Croton04} 
Croton D.J., \etal, 2004, MNRAS 352, 1232
\bibitem [\protect\citename{Croton \etal} 2007]{Croton07} 
Croton D.J., \etal, 2007, MNRAS 379, 1562

\bibitem [\protect\citename{Driver \etal} 2009]{Driver09}
Driver S.P., \etal, 2009, A\&G, 50e, 12
\bibitem [\protect\citename{Driver \etal} 2010]{Driver11}
Driver S.P., \etal, 2011, MNRAS 413, 971

%{\bf
\bibitem [\protect\citename{Evrard \etal} 2002]{Evrard02} 
Evrard A., \etal, 2002, ApJ 573, 7 
%}
%\bibitem [\protect\citename{Gazta\~naga} 1994]{Gaztanaga94} 
%Gazta\~naga E., 1994, MNRAS, 268, 913 
\bibitem [\protect\citename{Gazta\~naga} 2005]{Gaztanaga05} 
Gazta\~naga E., Norberg P., Baugh C.M., Croton D.J., 2005, MNRAS 364, 620
\bibitem [\protect\citename{Gott \etal} 2005]{Gott05} 
Gott J.R., \etal, 2005, ApJ 624, 463

\bibitem [\protect\citename{XX} XX]{Hartlap07} 
Hartlap J., Simon P., Schneider P., 2007, A\&A 464, 399

%{\bf
\bibitem [\protect\citename{XX} XX]{McBride11a} 
McBride C.K., Connolly A.J., Gardner J.P., Scranton R., Newman J.A., Scoccimarro R., Zehavi I., Schneider D.P.,  2011, ApJ 726, 13
\bibitem [\protect\citename{XX} XX]{McBride11b} 
McBride C.K., Connolly A.J., Gardner J.P., Scranton R., Scoccimarro
R., Berlind A.A., Marin F., Schneider D.P.,  2011,  2010arXiv1012.3462
%}
%{\bf
\bibitem [\protect\citename{XX} XX]{Miller74} 
Miller R.G., 1974, Biometrika, 61, 1
%}
\bibitem [\protect\citename{Nichol \etal} 2006]{Nichol06} 
Nichol R.C, \etal, 2006, MNRAS 368, 1507
\bibitem [\protect\citename{XX} XX]{Norberg02} 
Norberg P., \etal, 2002, MNRAS 332, 827
\bibitem [\protect\citename{XX} XX]{Norberg09}
Norberg P., \etal, 2009, MNRAS 396, 19

%{\bf
\bibitem [\protect\citename{XX} XX]{Roche93}
Roche N., Shanks T., Metcalfe N., Fong R., 1993, MNRAS 263, 360
\bibitem [\protect\citename{XX} XX]{Roche99}
Roche N., \& Eales S.A., 1999, MNRAS 307, 703
%}

\bibitem [\protect\citename{XX} XX]{Strauss02}
Strauss M.A., \etal, 2002, AJ, 124, 1810

\bibitem [\protect\citename{XX} XX]{Tucker06} 
Tucker D.L., \etal, 2006, AN, 327,821 
%{\bf
\bibitem [\protect\citename{XX} XX]{Turkey58} 
Tukey J.W., 1958, Ann. Math. Stat. 29, 614 
%}
\bibitem [\protect\citename{XX} XX]{Yaryura10} 
Yaryura Y., Baugh C.M., \& Angulo R.,  2011, MNRAS 413, 1311
\bibitem [\protect\citename{York \etal} 2000]{York00} 
York D.G., \etal, AJ, 120, 1579, 2000 

%{\bf
\bibitem [\protect\citename{XX} XX]{Zehavi02} 
Zehavi I., \etal, 2002, ApJ 571, 172
\bibitem [\protect\citename{XX} XX]{Zehavi04} 
Zehavi I., \etal, 2004, ApJ 608, 16
\bibitem [\protect\citename{XX} XX]{Zehavi05} 
Zehavi I., \etal, 2005, ApJ 630, 1
\bibitem [\protect\citename{XX} XX]{Zehavi11} 
Zehavi I., \etal, 2011, 2010arXiv1005.2413
%}

% Useful link for further references on basic statistics:
% http://elibrary.unm.edu/sora/Auk/v104n01/p0144-p0146.html

\end{thebibliography}
\end{document}